%% file: main.tex
\documentclass[11pt,a4paper]{article}

\usepackage{ifthen} 
\newboolean{pdflatex}
\setboolean{pdflatex}{true} 

\newboolean{articletitles}
\setboolean{articletitles}{true} 

\newboolean{uprightparticles}
\setboolean{uprightparticles}{false} 

\newboolean{inbibliography}
\setboolean{inbibliography}{false} 

\input{jinst_preamble}

\input{00_testbeam_symbols}

\usepackage{makecell}

\input{jinst_title}

\begin{document}
\maketitle


%


\input{01_introduction}

\input{02_TPA}

\input{03_Sensor_Timepix}

\input{04_Time_resolution_results}
\input{05_Conclusion}

\input{06_acknowledgements}

\clearpage
\addcontentsline{toc}{section}{References}
\setboolean{inbibliography}{true}
\bibliographystyle{JHEP}
\bibliography{main}

\end{document}

%% file: jinst_preamble.tex
\usepackage{microtype}
\usepackage{xspace} 
\usepackage{caption} 

\usepackage{graphicx}  
\usepackage{color}
\usepackage{colortbl}
\graphicspath{{./figs/}} 
\usepackage{pdflscape}

\usepackage{siunitx} 
\usepackage{amsmath} 
\usepackage{amssymb}
\usepackage{amsfonts}
\usepackage{upgreek} 

\usepackage{jinstpub}

\usepackage{natbib}
\usepackage{graphicx}
\usepackage{siunitx}
\usepackage{physics}
\usepackage{cleveref}
\usepackage{tikz}
\usepackage{tikz-3dplot}
\usetikzlibrary{arrows.meta}
\usepackage[nobiblatex]{xurl}

%% file: 00_testbeam_symbols.tex
\def\fig {figure~}
\def\figNoSpace {figure}
\def\Fig {Figure~}

\def\sect {section~}
\def\sectNoSpace {section}

\def\eq {Eq.~}

\def\np {\mbox{{n-on-p}}\xspace}
\def\pn {\mbox{{p-on-n}}\xspace}

\DeclareSIUnit\mneq{\cdot\text{1~{MeV~n}$_\text{eq}$\,cm$^{-2}$}\xspace}

\def\mum  {\ensuremath{{\,\upmu\mathrm{m}}}\xspace}

%% file: jinst_title.tex
\title{Charge and temporal characterisation of silicon sensors using a two-photon absorption laser}
\author[a,b,1]{R.~Geertsema\note{Corresponding author},}
\author[a]{K.~Akiba,}
\author[a]{M.~van Beuzekom,}
\author[a]{T.~Bischoff,}
\author[a]{K.~Heijhoff,}
\author[a,c]{H.~Snoek}

\affiliation[a]{Nikhef, Amsterdam, the Netherlands}
\affiliation[b]{Universiteit Maastricht, Maastricht, the Netherlands}
\affiliation[c]{Universiteit van Amsterdam, Amsterdam, the Netherlands}

\emailAdd{r.geertsema@nikhef.nl}

\abstract{

First measurements are presented from a newly commissioned two-photon absorption (TPA) setup at Nikhef. The characterisation of the various components of the system is discussed. Two planar silicon sensors, one being electron collecting and one hole collecting, are characterised with detailed measurements of the charge collection and time resolution. The TPA spot is determined to have a radius of \SI{0.975(11)}{\micro\meter} and length of \SI{23.8}{\micro\meter} in silicon. The trigger time resolution of the system is shown to be maximally \SI{30.4}{\pico\second}. For both sensors, uniform charge collection is observed over the pixels, and the pixel side metallisation is imaged directly using the TPA technique. The best time resolution for a single pixel is found to be \SI{600}{\pico\second} and \SI{560}{\pico\second} for the electron and hole collecting sensors respectively, and is dominated by ASIC contributions. Further scans at different depths in the sensor and positions within the pixels have been performed and show a uniform response. It is concluded that the TPA setup is a powerful tool to investigate the charge collection and temporal properties of silicon sensors.

}

\keywords{Solid state detectors; Particle tracking detectors (Solid-state detectors); Hybrid detectors; Timing detectors}


%% file: 01_introduction.tex
\section{Introduction}

Single-photon absorption (SPA) techniques have been extensively used to characterise silicon sensors, as reported in~\cite{eremin1996development}. These techniques are however limited due to the nature of photon interactions in silicon at the wavelengths typically used, resulting in small absorption depths and thus most charge is liberated close to the surface and not deep into the bulk. In contrast two-photon absorption (TPA) can access regions deeper into the bulk of sensors by using photons with energies below the silicon bandgap energy. 
Recently TPA techniques have seen an increase in their application for the characterisation of silicon detectors in high-energy particle physics~\cite{garcia2020high, ugobono2018sissa}, and initial spatial measurements indicate that TPA can be used to probe the silicon bulk with high resolving power~\cite{garcia2020high}.

To achieve the intensities required for TPA to occur without damaging the material, the optical pulses need to have a very short time duration. These pulses therefore also provide the possibility of sub-picosecond time resolution, since all charge carriers are typically liberated in a time window of hundreds of femtoseconds. Because of the short pulses, TPA systems can also be employed to study the time resolution of silicon detectors with high temporal precision. In this paper an overview of the TPA setup commissioned at Nikhef is outlined, and the time resolution of the system is determined. The system has been used to characterise the charge collection spatially as well as the time resolution of planar silicon sensors bump bonded to Timepix3 application-specific integrated circuits (ASIC)~\cite{poikela2014timepix3}. 

The paper is organised as follows: the TPA system is outlined in \sect\ref{sec:TPA} along with the characterisation studies of this system that have been performed. The detectors that have been used to perform the first characterisations are discussed in \sect\ref{sec:Sensor}. The charge collection and temporal measurements are described in \sect\ref{sec:Measurements}, followed by the conclusions in \sect\ref{sec:Conclusion}.

%% file: 02_TPA.tex
\section{Two-photon absorption laser system}
\label{sec:TPA}

Typically, low energy ($\sim$ $\mathcal{O}\left(1 \right)$ eV) photons propagating in a semiconductor are absorbed via a single interaction with an electron. The energy of the photon may excite an electron from the valence band of the semiconductor to the conduction band. This excitation can only occur if the initial photon energy is larger than the bandgap energy of the semiconductor. This excitation can also be achieved by a simultaneous interaction of multiple photons, if their combined energy is high enough to bridge the bandgap and if the interaction of the later photon(s) occurs within the de-excitation time of the electron. This process is called multi-photon absorption, and more specifically, the excitation that requires two photons is called two-photon absorption (TPA). The multi-photon process requires a high density of photons to achieve a significant probability of multiple photons interacting simultaneously. As described in~\cite{wiehe2020development}, the optical absorption in silicon is given as a function of depth $z$ along the beam propagation direction:

\begin{equation}
    \frac{\text{d}I}{\text{d}z}=-\alpha I -\sigma_{ex} N I -  \beta_2 I^2 + \mathcal{O}\left( I^3 \right),
    \label{eq:absorption}
\end{equation}
where $I$ is the intensity, $\alpha$ and $\beta_2$ are the SPA and TPA absorption coefficients, respectively, $\sigma_{ex}$ is the cross section for free-carrier absorption and $N$ the number of free charge carriers. From \cref{eq:absorption} it can be seen that the number of charge carriers created using SPA scales linearly with the intensity and quadratically for TPA.

\begin{figure}[h!]
    \centering
  {\includegraphics[width=0.49\textwidth]{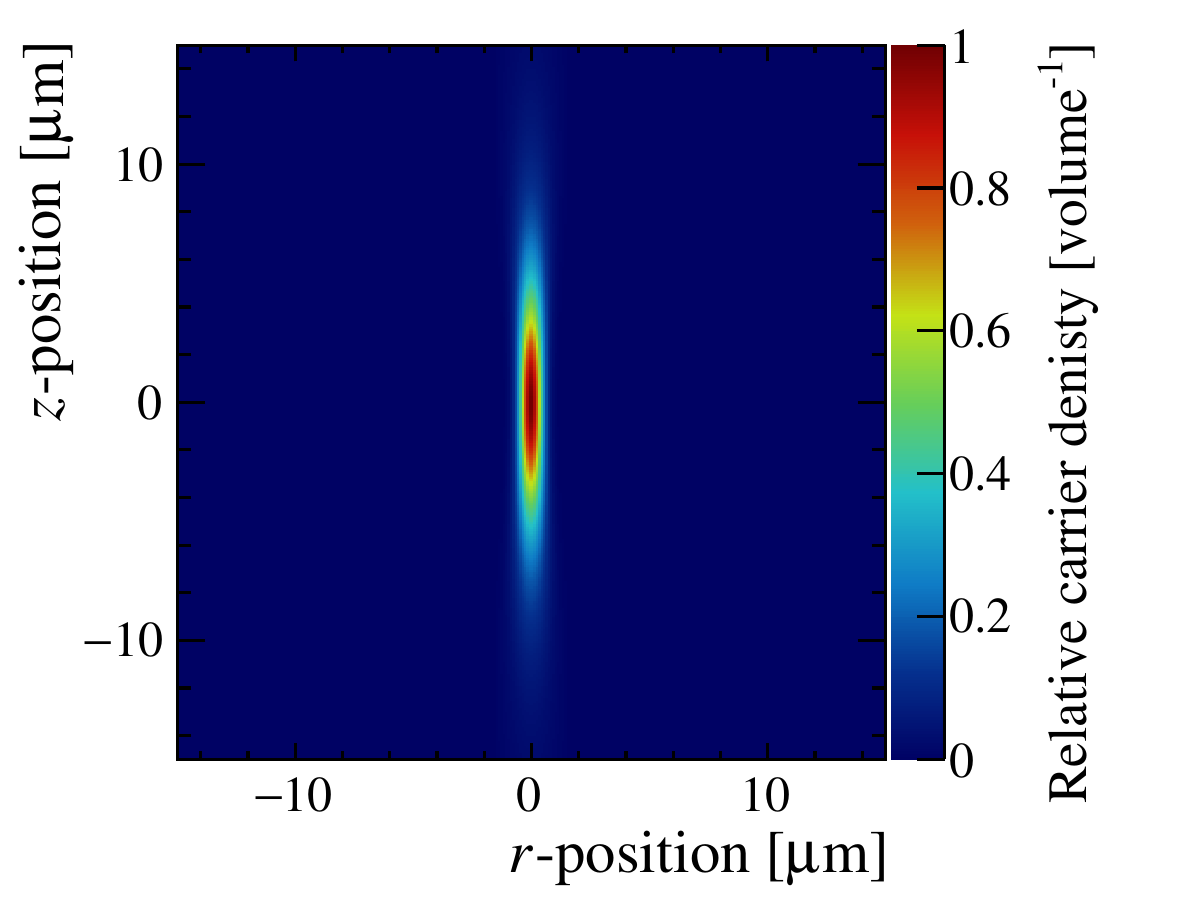}}
  \caption{The relative charge carrier density that is liberated along the beam path. This profile shows the intensity squared from \eq\ref{eq:intensity}. 
  }
  \label{fig:TPAspot}
\end{figure}

\begin{figure}[h!]
    \centering
  { 
  \usetikzlibrary{3d,spy}
\tdplotsetmaincoords{75}{65}
\begin{tikzpicture}[scale=.006in,tdplot_main_coords,spy using outlines={rectangle, magnification=4, size=2cm, connect spies}]
\tdplotsetrotatedcoords{90}{0}{0}  
\begin{scope}[tdplot_rotated_coords]
\draw[thick,-latex] (0,10,0) -- (1.5,10,0) node[below] {$x$};
\draw[thick,-latex] (0,10,0) -- (0,11.5,0) node[below] {$y$};
\draw[thick,-latex] (0,10,0) -- (0,10,1.5) node[right] {$z$};

\pgfmathsetmacro{\cubex}{5}
\pgfmathsetmacro{\cubey}{5}
\pgfmathsetmacro{\cubez}{1.5}
\pgfmathsetmacro{\cubezoffset}{-2}
\draw[black] (0,0,\cubezoffset) -- ++(\cubex,0,0) -- ++ (0,\cubey,0) -- ++ (-\cubex,0,0)  -- cycle;
\draw[black] (0,0,\cubezoffset) -- ++(0,0,\cubez) -- ++ (\cubex,0,0) -- ++ (0,0,-\cubez);
\draw[black] (0,0,\cubez+\cubezoffset) -- ++(0,\cubey,0) -- ++ (0,0,-\cubez);
\draw[black] (0,\cubey,\cubez+\cubezoffset) -- ++(\cubex,0,0) -- ++ (0,-\cubey,0);
\draw[black] (\cubex,\cubey,\cubez+\cubezoffset) -- ++(0,0,-\cubez);

\pgfmathsetmacro{\cubex}{5}
\pgfmathsetmacro{\cubey}{5}
\pgfmathsetmacro{\cubez}{10}
\draw[black,fill=white] (0,0,0) -- ++(\cubex,0,0) -- ++ (0,\cubey,0) -- ++ (-\cubex,0,0)  -- cycle;
\draw[black] (0,0,0) -- ++(0,0,\cubez) -- ++ (\cubex,0,0) -- ++ (0,0,-\cubez);
\draw[black] (0,0,\cubez) -- ++(0,\cubey,0) -- ++ (0,0,-\cubez);
\draw[black] (0,\cubey,\cubez) -- ++(\cubex,0,0) -- ++ (0,-\cubey,0);
\draw[black] (\cubex,\cubey,\cubez) -- ++(0,0,-\cubez);

\draw   (1,1.78) .. controls (1,1.35) and (1.35,1) .. (1.78,1) -- (3.22,1) .. controls (3.65,1) and (4,1.35) .. (4,1.78) -- (4,3.22) .. controls (4,3.65) and (3.65,4) .. (3.22,4) -- (1.78,4) .. controls (1.35,4) and (1,3.65) .. (1,3.22) -- cycle ;

\begin{scope}[canvas is xz plane at y=0,transform shape,every node/.style={scale=2}]
\node [align=center] (Text) at (1.3,9.4) {Sensor};
\end{scope}

\begin{scope}[canvas is xz plane at y=0,transform shape,every node/.style={scale=2}]
\node [align=center] (Text) at (1.3,-1.25) {ASIC};
\end{scope}

\draw[line width=0.05mm]   (.80,0,7.50) .. controls (.80,0,7.0858) and (.8895,0,6.75) .. (1.00,0,6.75) .. controls (1.1105,0,6.75) and (1.20,0,7.0858) .. (1.20,0,7.50) .. controls (1.20,0,7.9142) and (1.1105,0,8.25) .. (1.00,0,8.25) .. controls (.8895,0,8.25) and (.80,0,7.9142) .. (.80,0,7.50) -- cycle ;

\draw[line width=0.05mm, {Latex[length=0.2mm, width=0.3mm]}-{Latex[length=0.2mm, width=0.3mm]}] (1,0,7.5) -- (1.15,-0.1,7.505);
\draw[line width=0.05mm, {Latex[length=0.2mm, width=0.3mm]}-{Latex[length=0.2mm, width=0.3mm]}] (1,0,7.5) -- (1.,0,8.25);

\spy [black,dashed, magnification=5.8,width=3cm,height=4.2cm] on (1.,0,3.55)  in node at (27,0,11.0);

\draw[thick,-latex] (14.3,0,-1.) -- (15.1,-0.4,-1) node[below] {$r$};
\draw[thick,-latex] (14.3,0,-1) -- (14.3,0,0) node[left] {$z$};

\node[text width=1cm] at (13.4,0,2.1) {$w_0$};
\node[text width=1cm] at (12.5,0,4.9) {$z_0$};
\begin{scope}[every node/.style={scale=.75}]
\node[text width=3cm] at (11.84,0,7.72) {TPA voxel};
\node[text width=3cm] at (8.3,0,10.22) {Back};
\node[text width=3cm] at (8.3,0,0.5) {Front};
\end{scope}

\end{scope}
\end{tikzpicture}

  }
  \caption{ Schematic overview of the sensor in which the TPA voxel is positioned. The definition of the coordinate system is indicated for both the sensor and the TPA voxel. The Timepix3 ASIC is also indicated.
  }
  \label{fig:dimensions}
\end{figure}
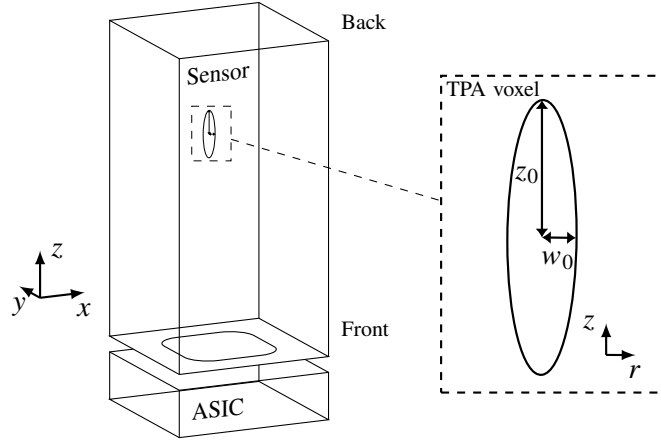

For a focused Gaussian beam, the intensity required for TPA to occur is only achieved close to the focal point (or by a very large beam power), thereby effectively constraining the TPA in space.
The focused beam creates an ellipsoidal volume (voxel - see \fig\ref{fig:TPAspot}) in which the liberation of charge carriers occurs. It is straightforward to scan the detector volume by moving it with respect to this voxel, making it possible to study the sensors in all three dimensions. The axis system that is used is indicated in \fig\ref{fig:dimensions}.

\begin{figure}[h!]
    \centering
  {\includegraphics[width=0.49\textwidth]{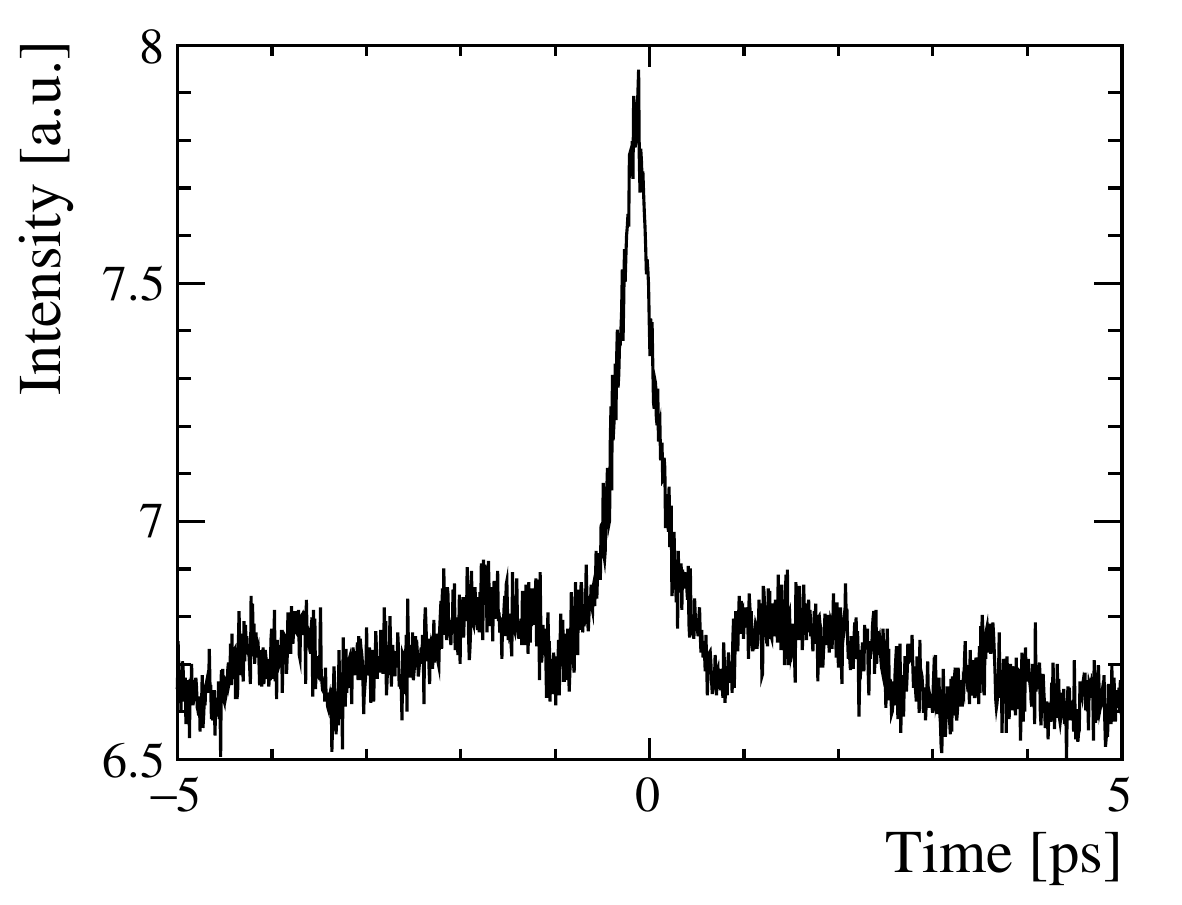}}\hspace{1mm}
  {\includegraphics[width=0.49\textwidth]{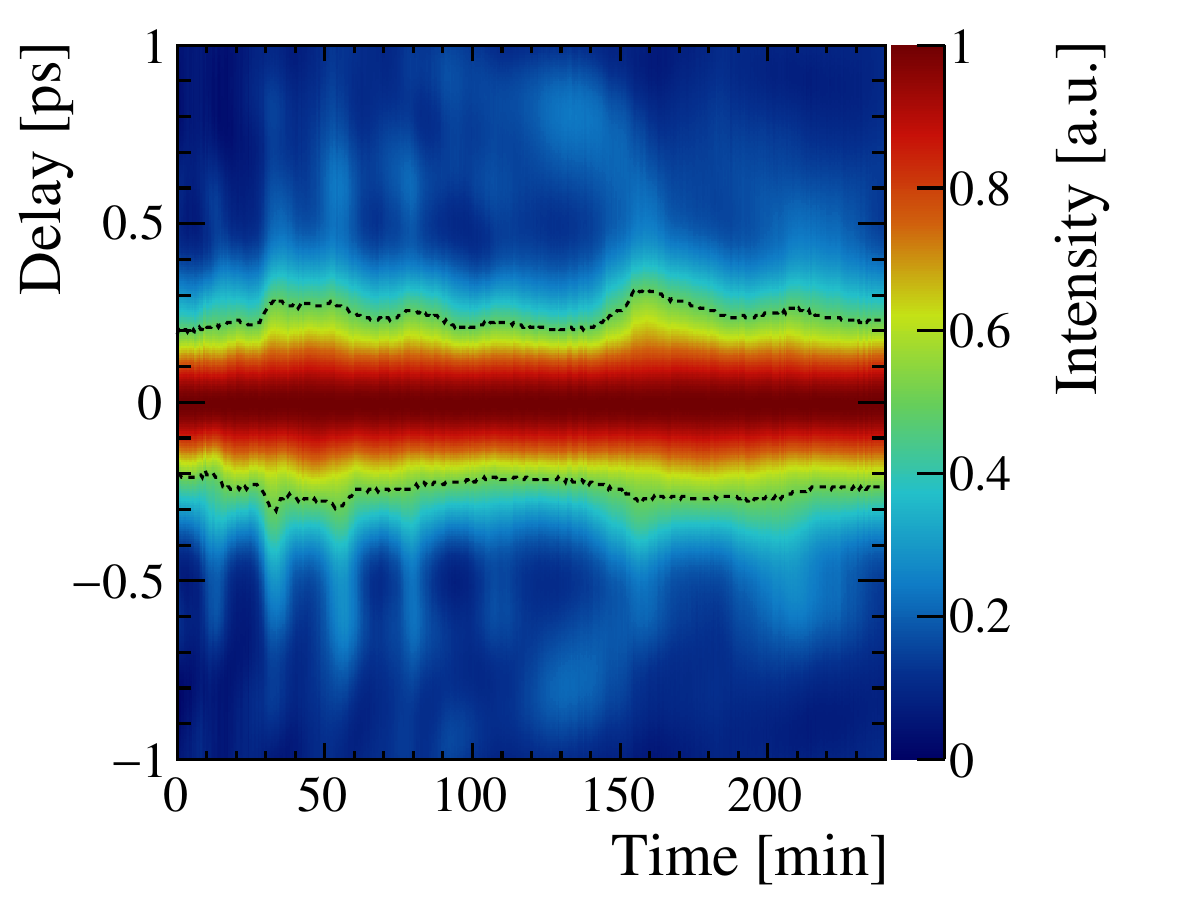}}
  \caption{ Left: a typical auto-correlation trace for a pulse length of \SI{380}{\femto\second}. Right: the pulse shape monitored using an autocorrelator over a time span of four hours. The dashed line indicates the FWHM of the pulses. 
  }
  \label{fig:temporalProfile}
\end{figure}

\begin{figure}[h!]
    \centering
  
  {\includegraphics[width=0.75\textwidth]{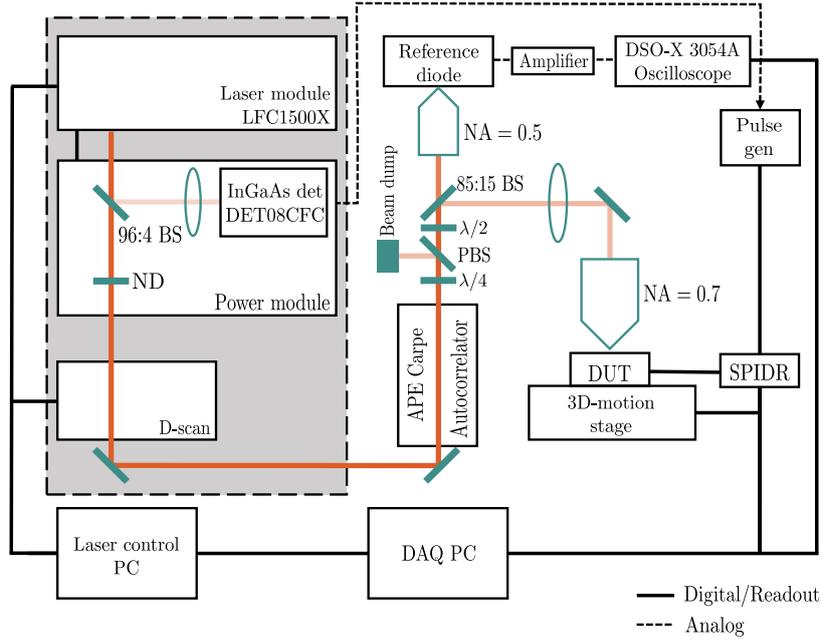}}
  \caption{Schematic overview of the TPA setup. The grey area indicates the section of the setup manufactured by Fyla. The orange line indicates the path of the laser beam, the dashed blue lines indicate all analogue connections, and the black lines all digital connections. The green components represent the optical components. 
  }
  \label{fig:TPAlayout}
\end{figure}

The system presented in this paper consists of a fiber based \SI{1550}{\nano\metre} pulsed laser system (LFC1500X) from Fyla\footnote{FYLA, Ronda Guglielmo Marconi 14, Parque Tecnológico 46980
Paterna, Valencia, Spain}. For all measurements it is operated at a repetition rate of \SI{15.92}{\kilo\hertz}, and a variable Neutral Density (ND) filter is used to tune the pulse energy. 
The Full Width at Half Maximum (FWHM) pulse width is measured with an APE Carpe autocorrelator~\cite{APECarpe} and varies between \SI{380}{\femto\second} and \SI{540}{\femto\second} throughout the measurements. A typical auto-correlation trace can be seen in \fig\ref{fig:temporalProfile}~(left). The FWHM can be tuned to larger values by changing the group delay dispersion (GDD) using a D-Scan 1.5 system\footnote{Sphere Ultrafast Photonics, Rua do Campo Alegre, n.º 1021, Edifício FC6
4169-007 Porto, Portugal}. A schematic overview of the system is shown in \fig\ref{fig:TPAlayout}. The laser system is controlled via a dedicated computer that can communicate with the computer (DAQ) that acquires the data of the Timepix3 ASIC. 

\begin{figure}
\centering
\begin{minipage}{.47\textwidth}
\centering
\includegraphics[width=1\linewidth]{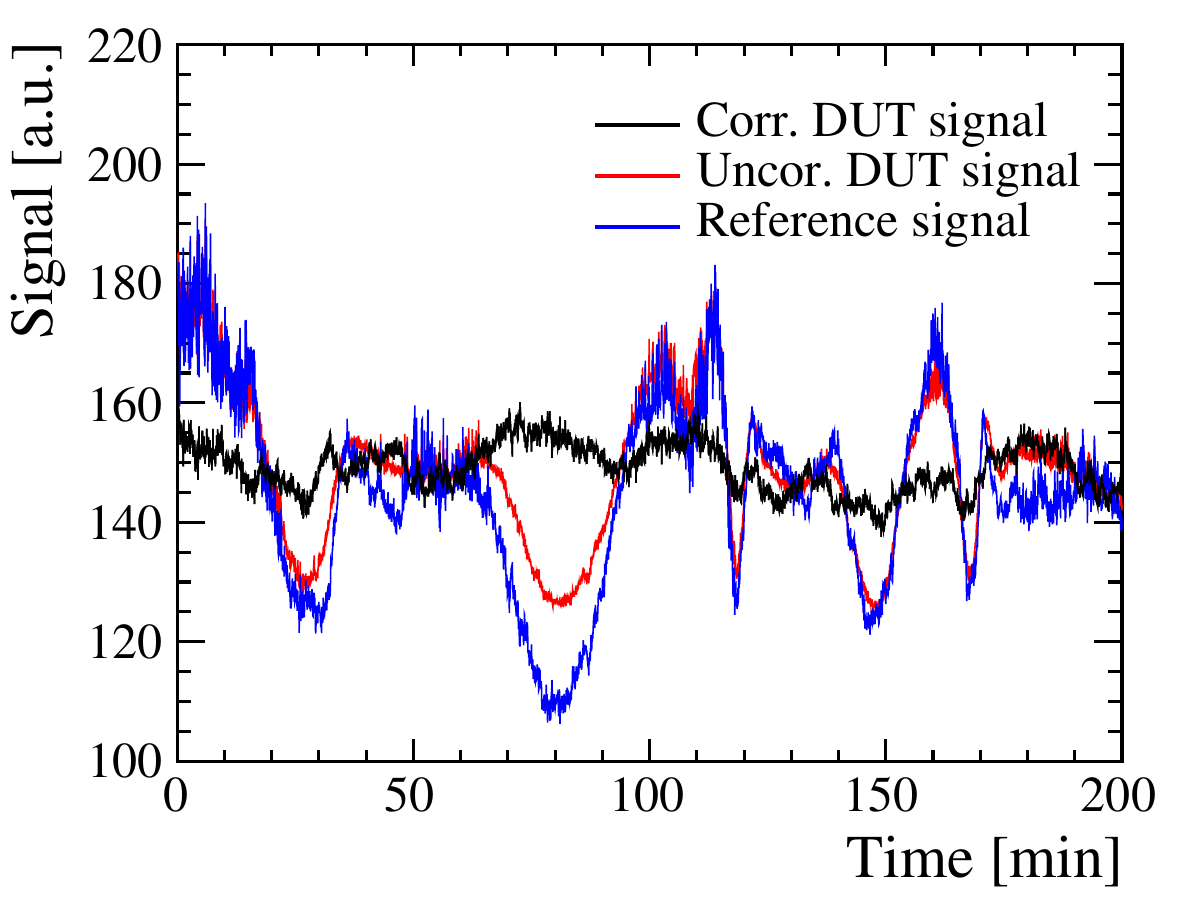}
\end{minipage}\qquad
\begin{minipage}{.47\textwidth}
\centering
\includegraphics[width=1\linewidth]{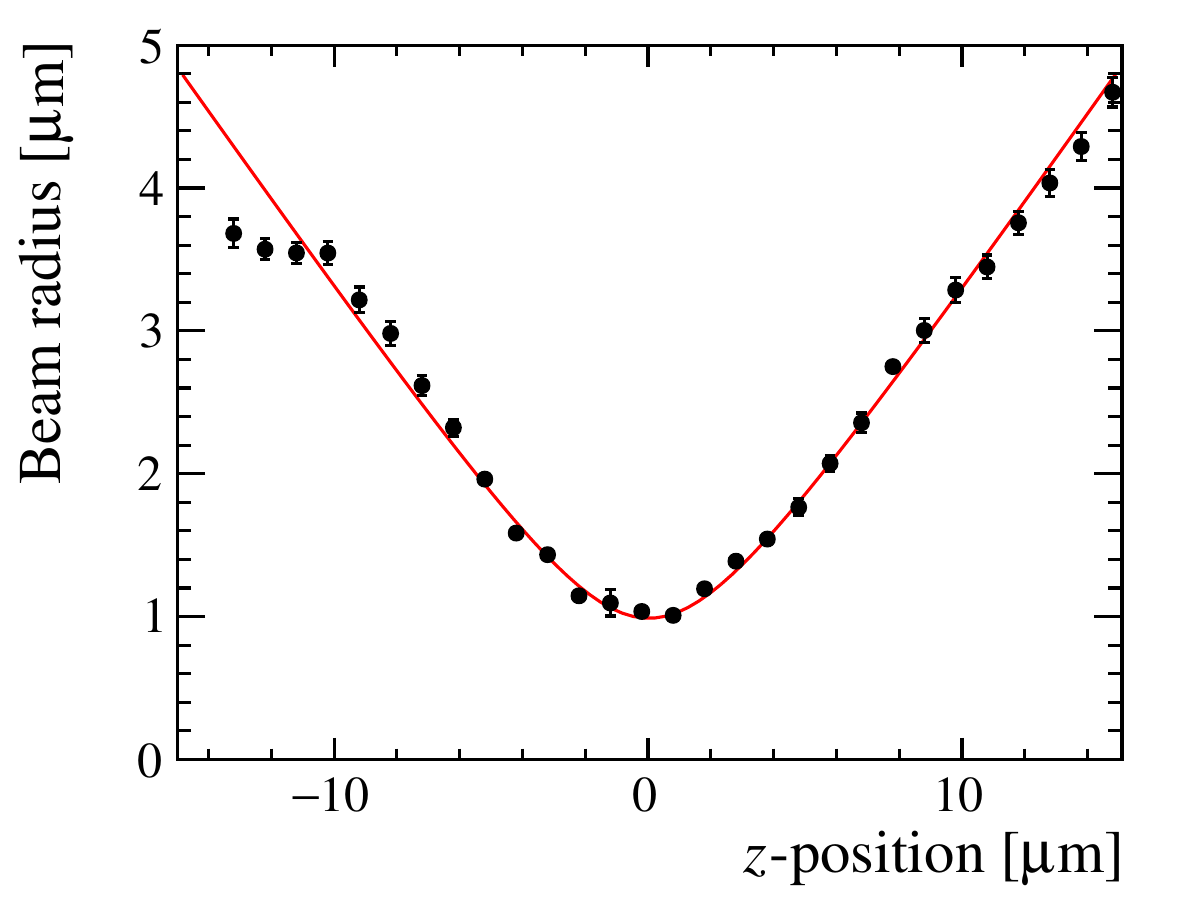}
\end{minipage}


\begin{minipage}[t]{.47\textwidth}
\centering
\captionsetup{width=.99\linewidth}
\caption{The uncorrected TPA signal of the DUT (red), the scaled TPA signal measured at the reference diode (blue) and the corrected TPA signal of the DUT (black). After correction a standard deviation of 1.6\% is achieved.}
\label{fig:correctionDiode}
\end{minipage}\qquad
\begin{minipage}[t]{.47\textwidth}
\centering
\captionsetup{width=.99\linewidth}

\captionof{figure}{The beam radius of the laser pulses as a function of height around the focal point. The black dots indicate the measured data and the red line indicates the fit. The first three data points are excluded from the fit.}
\label{fig:beamWaist}
\end{minipage}
\end{figure}

The pulse width is observed to vary over time for a constant dispersion compensation. To characterise these fluctuations, the pulse width is monitored over four hours, showing a peak-to-peak variation of \SI{150}{\femto\second}. This data can be seen in \fig\ref{fig:temporalProfile}~(right). Along with a small ($<1\%$) variation of the output power of the laser, this results in an undesired variation of the number of charge carriers liberated in the semiconductor per laser pulse over time. 
To correct for this variation offline, a fraction of the beam is diverted to a silicon diode. 
The same material, silicon, is chosen to have a direct correspondence to TPA effects since it is inherently nonlinear. 
At the present time the calibration fraction is about 85\% of the total laser power as it also serves to decrease the incident power onto the Detectors Under Test (DUT). 
The output of this reference diode is first amplified with a Cividec C2-TCT amplifier~\cite{Cividec}. 
This amplified signal is monitored using an oscilloscope (DSO-X 3054A) which integrates the signal of each pulse. 
A quarter-wave and half-wave plate combined with a polarising beam splitter are placed after the autocorrelator to ensure constant polarisation at both the reference diode and the DUT. 
The half-wave plate is used to rotate the polarisation such that the maximum transmission at the air-silicon interface is achieved. The resulting signal fluctuation after this correction is 1.6\% (standard deviation) of the signal; the data are shown in \fig\ref{fig:correctionDiode}.

To minimise the voxel size, the laser beam is focused using a microscope objective. As described in \cite{svelto1998principles}, the temporal and spatial evolution of a Gaussian beam with a temporal width $\tau$ that is focused to a beam waist $w_0$ is given by~\cite{svelto1998principles}:
\begin{equation}
    I\left( r, z, t \right) = \frac{E_p}{\tau} \frac{4\sqrt{\ln{2}}}{\pi^{3/2}w^2\left( z \right)} \text{exp}\left( -\frac{2r^2}{w^2\left( z \right)} \right) \text{exp}\left( -4 \ln{2\frac{t^2}{\tau^2}} \right),
    \label{eq:intensity}
\end{equation}
where $E_p$ is the pulse energy. The two-sigma beam radius $w(z)$ of the electric field along the beam propagation direction is given by
\begin{equation}
    w\left( z \right) = w_0 \sqrt{1 + \left( \frac{\lambda z  }{\pi w_0^2 n}\right)^2}=w_0\sqrt{1+\left( \frac{z}{z_0} \right)^2},
    \label{eq:beamRadius}
\end{equation}
where $n$ is the refractive index of the material, $\lambda$ the wavelength, and $z_0$ the Rayleigh length ($z_0 = \pi w_0^2 n /\lambda$). Since the Rayleigh length is related to the beam waist size, the dimensions of the voxel are proportional to $w_0$. The voxel dimensions and axis system are shown in \fig\ref{fig:dimensions}. Since a small voxel is desired, an objective with a high magnification needs to be used. The objective chosen is a 100x Mitutoyo\footnote{Mitutoyo Corporation, 20-1, Sakado 1-Chome, Takatsu-ku, Kawasaki-shi, Kanagawa 213-8533, Japan} Plan Apo NIR HR Infinity Corrected ($\text{NA} = 0.7$) for tests of the thin sensors and a 100x Mitutoyo Plan Apo NIR Infinity Corrected ($\text{NA} = 0.5$) for tests of the thicker sensors. The lower NA is required for thicker sensors since the increased spherical aberration for the higher NA objective causes a TPA intensity loss that increases linearly with depth~\cite{gu2000advanced}. The beam radius of the \SI{1550}{\nano\meter} beam around the focal point has been measured using a knife-edge method, and the results are shown in \fig\ref{fig:beamWaist}. Fitting \eq\ref{eq:beamRadius} to the data yields a value of \SI{0.975(11)}{\micro\meter} for the minimum beam radius and a value of \SI{3.01(5)}{\micro\meter} for the Rayleigh length in air. For this fit the first three data points (from -14 to -\SI{10}{\micro\meter}) are excluded. These points deviate from the expected curve, however an explanation for this behaviour is not yet found.

The voxel becomes effectively longer in silicon due to the change in refractive index at the air-silicon interface, as described in \cite{wiehe2020development}. Therefore, the effective position $z'$ within the silicon is given by
\begin{equation}
    z' = z \sqrt{ \frac{ z_0 \pi n^3 }{ z_0 \pi n - \lambda n^2 +\lambda } }.
\end{equation}
This effect results in a larger Rayleigh length in silicon, and yields $z_{0,\text{Si}}=11.9$~\si{\micro\meter}. The width of the voxel, is however not affected by the refractive index, and thus remains the same as in air. This correction is applied to determine the actual position of the focal point within the silicon detector during the measurements. 

\subsection{Trigger system}

A trigger signal is created in the laser power module using a \SI{5}{\giga\hertz} InGaAs detector (Thorlabs DET08CFC~\cite{ThorlabsDet}) with a rise time of \SI{70}{\pico\second}. A beam splitter is used to redirect 4\% of the beam to this detector. 
The time resolution of this detector has been determined by measuring the remaining 96\% of the beam with a Thorlabs PDA05CF2~\cite{ThorlabsDet2}, which has a rise time of \SI{2.3}{\nano\second}. 
The time difference between the rising edge of the signals (using a threshold of 50\% of the peak height) of the two detectors was determined using an oscilloscope. The standard deviation of the time difference is taken as the resolution of the two detectors combined. With a statistical sample of 150k measurements, the combined resolution is \SI{43}{\pico\second}. 
Lower time resolutions are measured with shorter intervals due to intensity fluctuations described earlier in this \sectNoSpace. Assuming that both detectors contribute equally, the trigger resolution is $43/\sqrt{2}=\ $\SI{30.4}{\pico\second}. However, based on bandwidth considerations it is expected that the trigger time resolution is much better than that of the reference detector.
Nevertheless, a time resolution of \SI{30.4}{\pico\second} is sufficiently small not to influence the 
measurements presented in this paper. In order to use the signal of the InGaAs detector, the level of the signal needs to be adapted to match that of the input of the TDC on the readout board (SPIDR - see \sect\ref{sec:Timepix3ASIC}). A pulse generator was used to adapt the level of the trigger signal.

%% file: 03_Sensor_Timepix.tex
\section{Sensor and Timepix3 ASIC}
\label{sec:Sensor}

The studies presented in this paper are obtained with hybrid pixel detectors, with sensors bump-bonded to Timepix3 ASICs~\cite{poikela2014timepix3}, read out by a SPIDR~\cite{visser2015spidr} readout system. In this section, the Timepix3 ASIC along with SPIDR is briefly discussed, after which the sensors characterised in this study are described.

\subsection{Timepix3 ASIC and SPIDR readout system}
\label{sec:Timepix3ASIC}

The Timepix3 ASIC is developed by the Medipix3~\cite{ballabriga2018asic} collaboration. It has a matrix of 55$\times$\SI{55}{\micro\meter\squared} structures each with a charge sensitive amplifier as a first stage and a discriminating circuit capable of measuring both the Time-of-Arrival (ToA) and the Time-over-Threshold (ToT) simultaneously for every signal. The time-to-digital converter (TDC) digitises the ToA using a \SI{640}{\mega\hertz} clock, giving a bin size of \SI{1.56}{\nano\second} (corresponding to a time resolution of \SI{451}{\pico\second}). The time resolution of the analogue front-end of Timepix3 has been determined to be approximately \SI{240}{\pico\second}~\cite{Laurens2021} for a typical signal amplitude and planar sensor. Variations in the clock distribution and in the manufacturing process contribute to a larger time resolution obtained for all pixels on average~\cite{heijhoff2020timing}, even excluding the contributions arising from the sensors. For all measurements in this paper, the Timepix3 is operated at a threshold of 1000~e$^-$, corresponding to approximately 5\% of the typical signal amplitude of a Minimum Ionising particle (MIP). The ToT information can be converted into units of electrons via a charge calibration process, as described in \cite{akiba2019lhcb}.

\begin{figure}[h!]
    \centering
  {\includegraphics[width=0.49\textwidth]{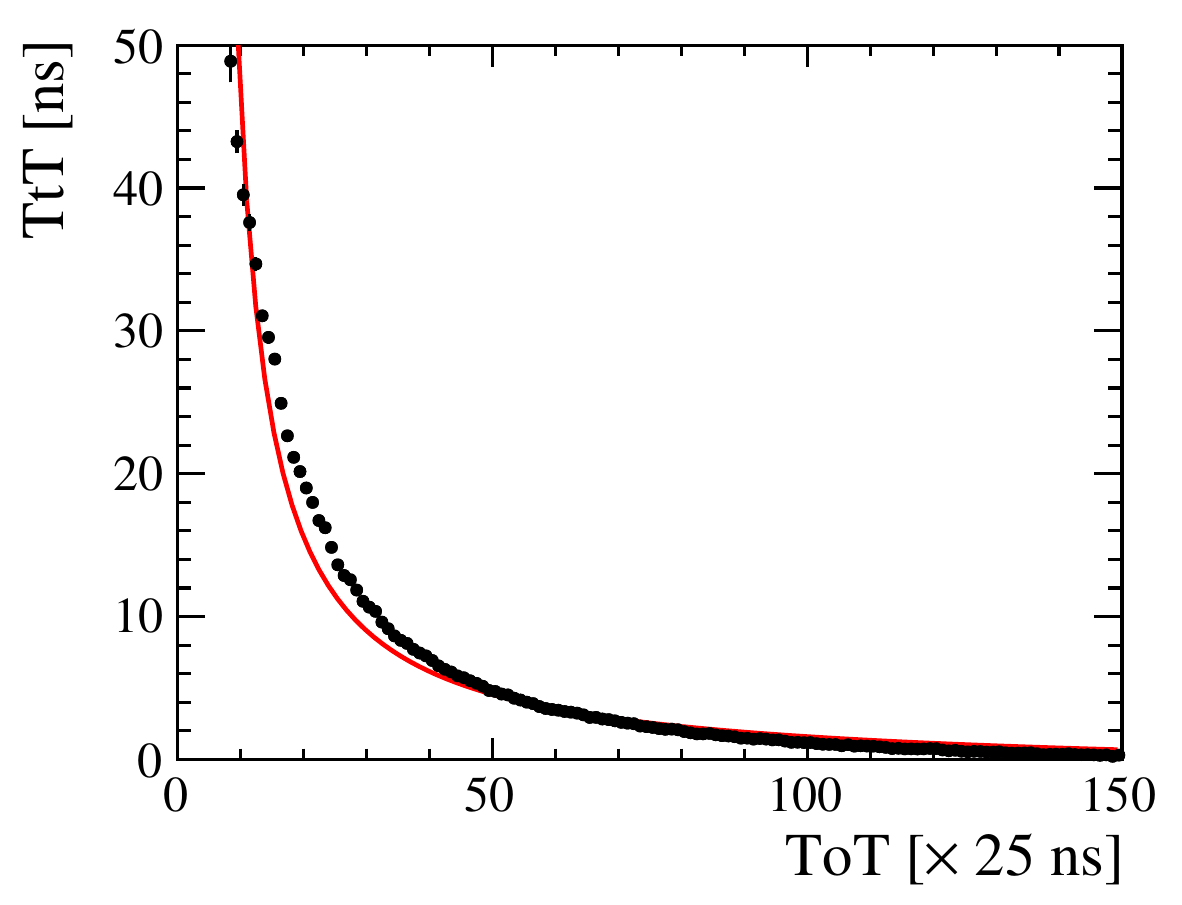}}\hspace{1mm}
  {\includegraphics[width=0.49\textwidth]{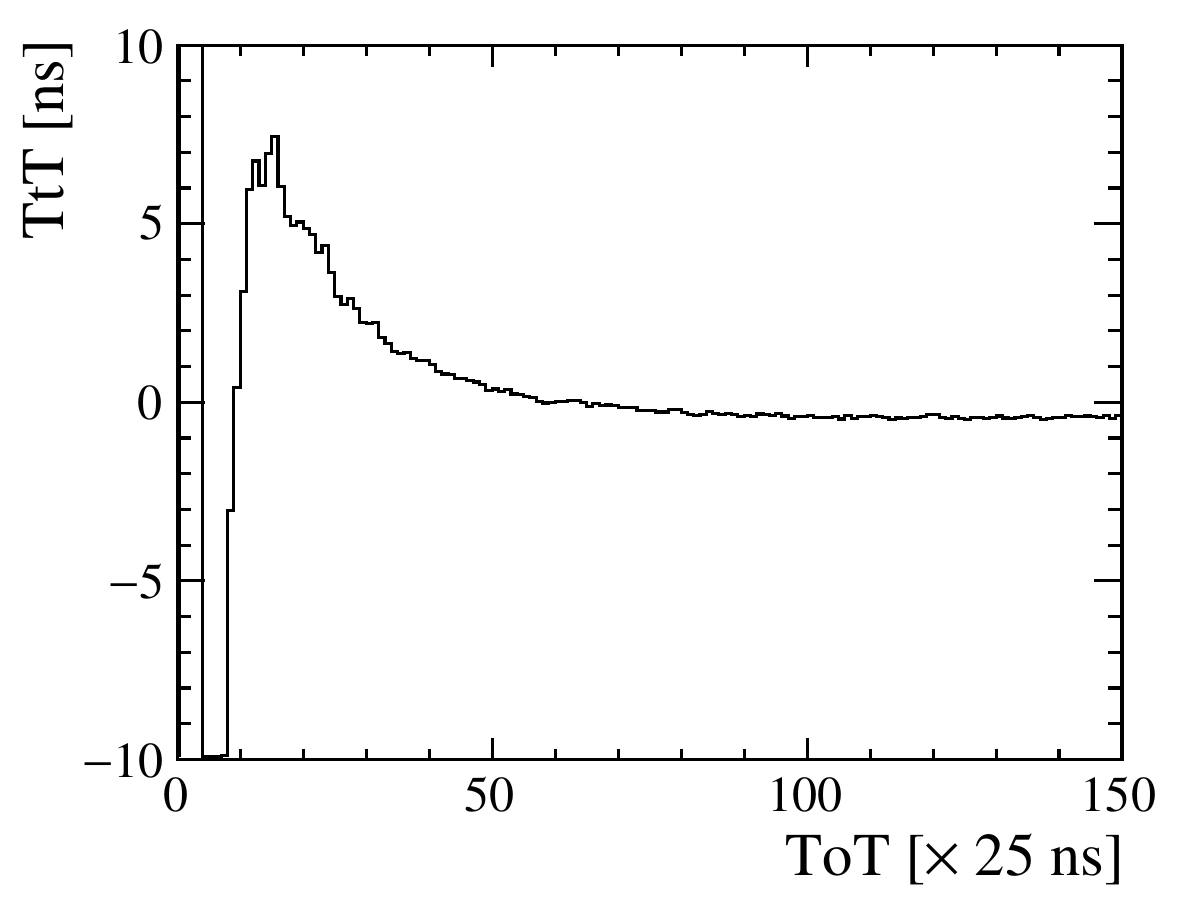}}
  \caption{ Left: a typical timewalk curve indicated in black, and the resulting parametrised correction is shown in red. Right: the TtT after applying the parametrised correction, for low ToT a systematic under- and over-correction can be observed.
  }
  \label{fig:Timewalk}
\end{figure}

Lower charge hits suffer from a lower induced current and thus cross the threshold level with a larger delay. The resulting delay (also known as timewalk) can be corrected for using the ToT information of each hit. The timewalk curve is parameterised as
\begin{equation}
    t\left( q \right) = \frac{A}{q-q_0}+C,
    \label{eq:timewalk}
\end{equation}
where $t$ is the timewalk, $q$ the charge, $q_0$ the charge corresponding to the onset of the asymptote, $A$ the slope, and $C$ the offset. A typical timewalk curve in shown in \fig\ref{fig:Timewalk}~(left). 
In this figure the black points indicate the measured data, and the line indicates the fit with \eq\ref{eq:timewalk}.
To parametrise the timewalk curve, the TPA voxel is positioned at the centre of the sensor implant, and the intensity of the laser is varied by rotating the variable ND filter. Using this method, the Time-to-Threshold (TtT) can be determined for a range of ToT (charge) values. The measured curve obtained from a fit to this data is used to correct for timewalk in the offline analysis. This parameterisation has been performed for every pixel individually. An example of the corrected TtT as a function of ToT is shown in \fig\ref{fig:Timewalk}~(right). A systematic under- and over-correction can be observed in the corrected TtT curve for low ToT values.

The Timepix3 is read out using a SPIDR readout system~\cite{visser2015spidr}. The SPIDR also has an on-board TDC with \SI{260}{\pico\second} bins based on a \SI{320}{\mega\hertz} clock generated using the \SI{40}{\mega\hertz} clock of the Timepix3. This TDC is used to register the trigger. In practise a non-uniformity on the TDC bin size is observed and the overall standard deviation of time measurements is \SI{77}{\pico\second}, instead of $260/\sqrt{12}=$~\SI{75}{\pico\second} for the ideal case.

In combination with the SPIDR TDC, the Timepix3 has a time resolution of $451\oplus240\oplus77=516$~ps (adding the separate contributions listed above in quadrature). This time resolution is insufficient for precise temporal measurements of the silicon sensor, but it may be used for the investigation of rough temporal characteristics of these sensors.

\subsection{Silicon sensors}
\label{sec:siliconSensors}
Two planar silicon pixel sensors are characterised for this paper. A 200\mum electron collecting \np sensor\footnote{Hamamatsu Photonics K. K., 325-6, Sunayama-cho, Naka-ku, Hamamatsu City, Shizuoka, 430-8587, Japan}, used in the prototyping phase of the LHCb VELO Upgrade R\&D, and a 300\mum hole-collecting \pn sensor\footnote{Canberra Semiconductors N.V., Lammerdries 25, 2259
Olen, Belgium}. 
These sensors have already been studied in detail in the past and therefore are ideal to test the TPA setup. From here on the side of the sensors with the segmented electrodes (to which the ASIC is connected) will be referred to as the front side. The other side with the common contact is referred to as the back side (see \fig\ref{fig:dimensions}). 

The time resolution of these assemblies (sensor and ASIC combined) can be approximated by a quadratic sum of three terms~\cite{Cartiglia_2014}: 
\begin{equation}
\label{eq:treso}
    \sigma_t^2 = \left( \left[ \frac{t_r V_{\text{th}}}{S} \right]_{\text{RMS}} \right)^2 + \left( \frac{t_r}{S/N} \right)^2 + \left( \frac{\text{TDC}_{\text{bin}}}{\sqrt{12}} \right)^2,
\end{equation}
where the terms make explicit the independent contributions from timewalk, jitter and TDC binning, respectively. Here $t_r$ is the rise time of the signal at the output of the amplifier, $V_{\text{th}}$ is the threshold of the discriminator, $S$ is the amplitude of the signal, $N$ is the noise of the front-end, and $\text{TDC}_{\text{bin}}$ is the TDC bin width. 

Since both an electron and a hole collecting device are studied, the charge collection times of the two are expected to be different. The Shockley-Ramo mechanism~\cite{shockley1938currents} contributes to the effective time resolution of the sensors since the drift velocity of holes in silicon is lower than that of electrons. The weighting field also slightly differs between the two sensors due to the difference in thickness and the small differences in the geometry of the pixel implants. Since these two effects will result in a difference in the induced current on the implants for the two sensors, the rise time (of the integrated current) of the electron collecting device is expected to be faster than that of the hole collecting device. These contributions, however, are small compared to ASIC contributions that have been discussed earlier. Therefore, the difference in time resolution of the two sensors that can be observed will be dominated by ASIC contributions and not by the difference in sensor type and thickness.

%% file: 04_Time_resolution_results.tex
\section{Measurements and results}
\label{sec:Measurements}

\subsection{Charge collection studies}
\label{sec:ChargeMeasurements}

\begin{figure}[t!]
  \centering
  {\includegraphics[width=0.49\textwidth]{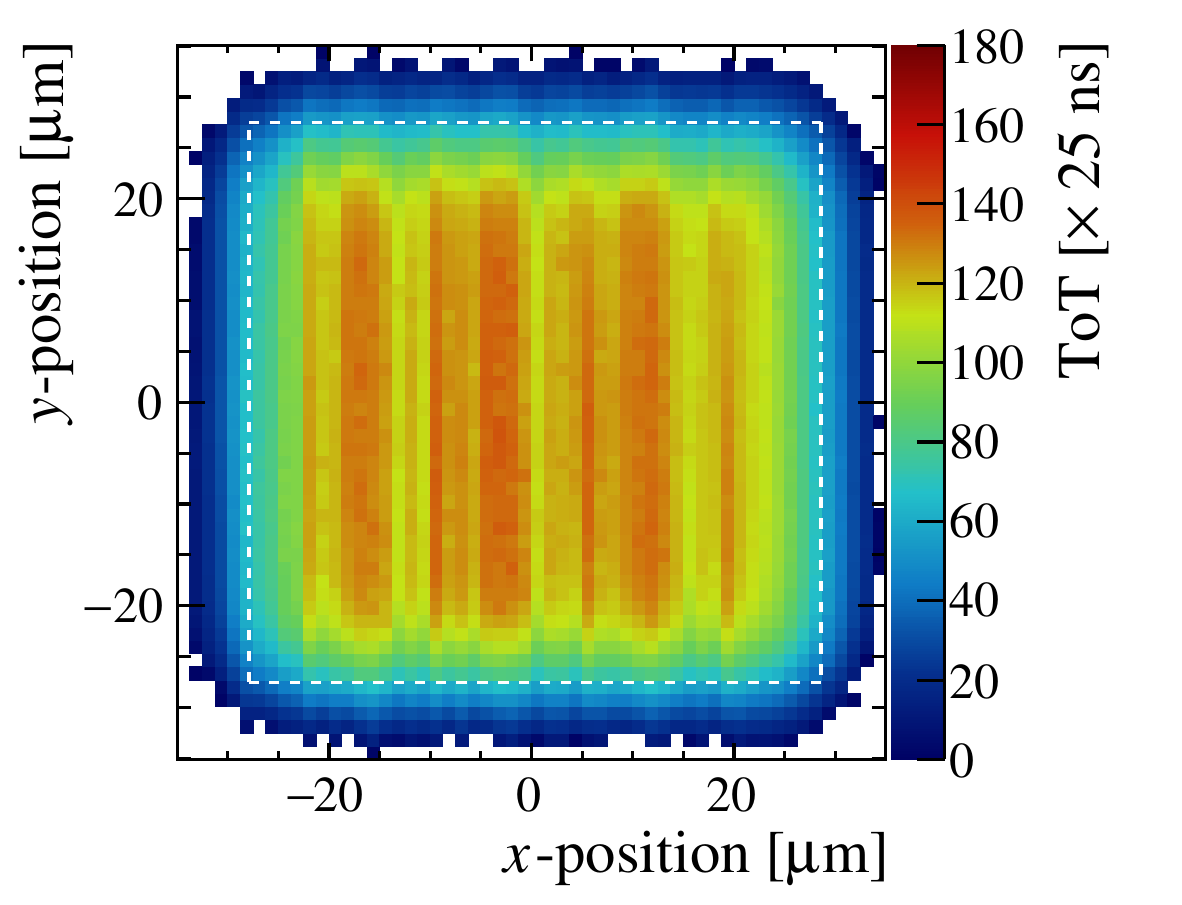}} \hspace{1mm}
  {\includegraphics[width=0.49\textwidth]{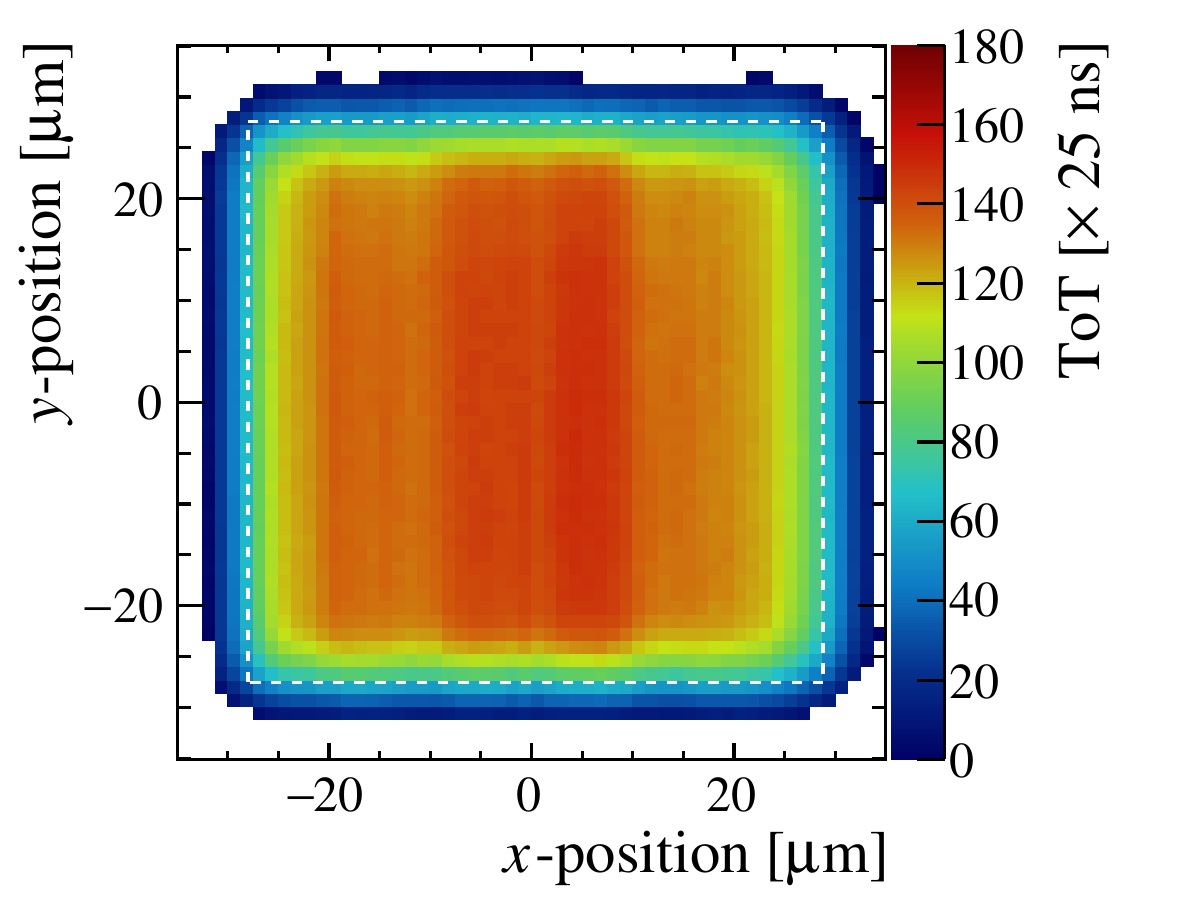}} \\
  {\includegraphics[width=0.49\textwidth]{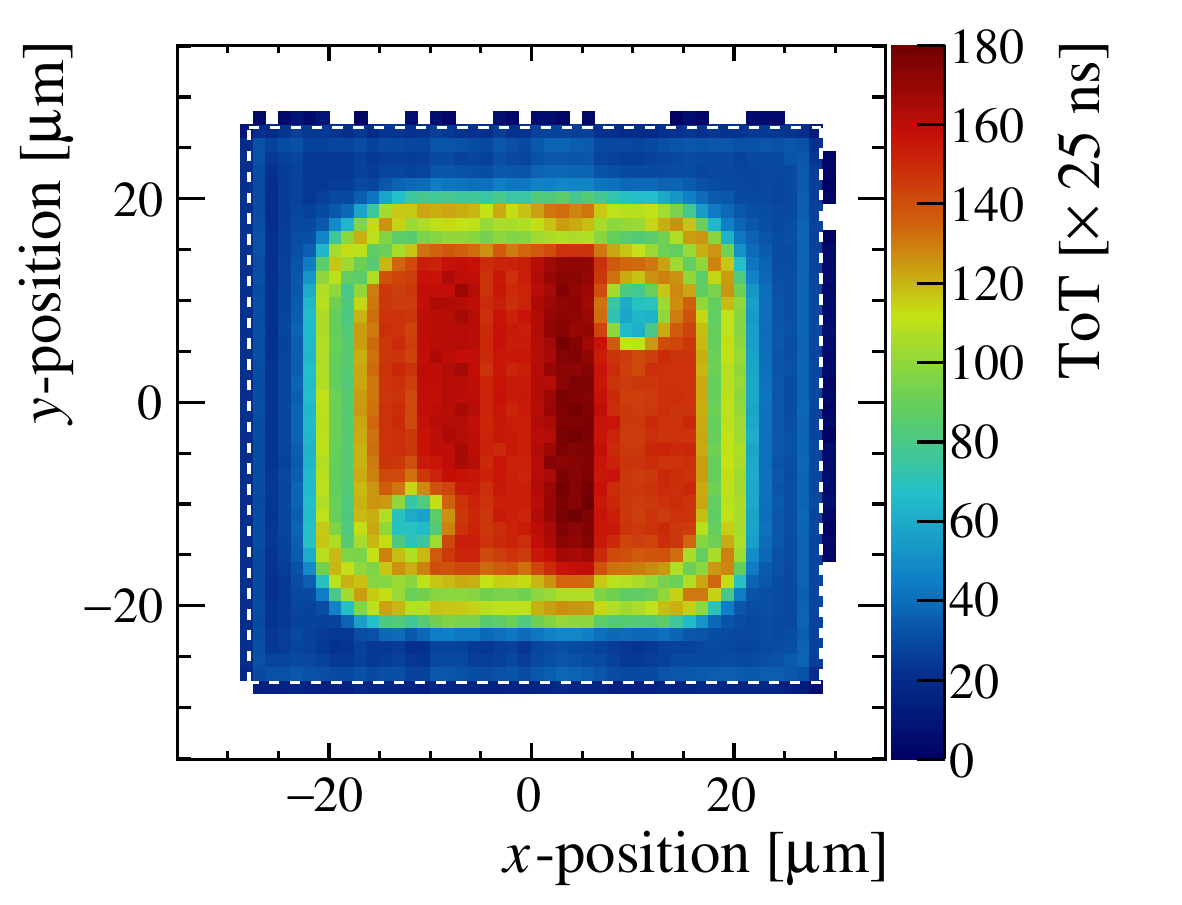}} \\
  \caption{Time-over-Threshold (ToT) scans of a pixel in the \np sensor, the dashed line indicates the area of the pixel. Top left: back side of sensor. Top right: middle of the sensor. Bottom: Front side of the sensor.}
  \label{fig:ToTscans}
\end{figure}

To study the charge collection properties of the sensors, the laser intensity is kept constant at a signal amplitude corresponding to a MIP, while the position of the TPA voxel is positioned at different places in the sensors. A single pixel of the \np sensor has been scanned at different depths in its bulk. These pixel scans are shown in \fig\ref{fig:ToTscans} for three different depths: back side (upper left plot), middle (upper right), and at the front side (lower). These scans extend \SI{27.5}{\micro\meter} into the neighbouring pixels (these pixels are not read out). Small fluctuations of the TPA intensity can be observed as vertical stripes in these scans, because the data were collected via a vertical raster scan. Therefore the time difference between two vertical lines is larger than between two neighbouring points at the same $x$-position (one single vertical line typically takes 10~minutes to measure, while a single point typically takes a few seconds). In all plots the boundary of the pixel is indicated by the dashed white lines. 

The scans of the back and middle of the sensor show uniform charge collection. The main difference between these two scans is the increased width of the transition region at the edge of the pixel. Charge generated at the back side of the sensor travels longer before being collected, and therefore also has larger lateral diffusion. The front-side measurement shows a structure due to the front-side metallisation. 
Owing to the strong reflection at the silicon-metal interface at the front of the sensor, the intensity of the laser in the silicon close to the interface is larger than without this reflection. This localised increased laser intensity results in an increase of TPA and thus in a larger deposited charge near the silicon-metal interface. 
This structure is a direct observation of the front-side metallisation as well as the angle of the metallisation with respect to the normal of the sensor. The position of two vias present in the pixel is observed, visible as two circles with lower ToT. A small increase in the ToT is observed at the implant border, indicating a feature on the front side which is unknown to us. It should be noted that the intensity correction implemented, is not able to fully correct the fluctuations near the metallisation since the reflected beam no longer is linearly proportional to the TPA signal in the reference diode. Since the thickness of the silicon part of the sensor is slightly larger at the pixel implant, and the alignment of the $z$-axis is performed at the pixel implant as well, the beam waist is outside of the bulk in the corners of the pixel. Therefore, relatively little charge is liberated at these positions compared to the implant. The same study was also performed for the \pn sensor. Besides a different front side metallisation, the results are similar and therefore not included here. 

The pixel width along the $x$-coordinate is determined with pixel scans to be \SI{55}{\micro\meter} without any additional calibration of the stage, however the width along $y$ is measured to be approximately \SI{2.4}{\micro\meter} smaller. After a calibration of the $y$-stage, this discrepancy has disappeared and the pixel widths along $y$ are also \SI{55}{\micro\meter}.

\begin{figure}[h!]
    \centering
  {\includegraphics[width=0.49\textwidth]{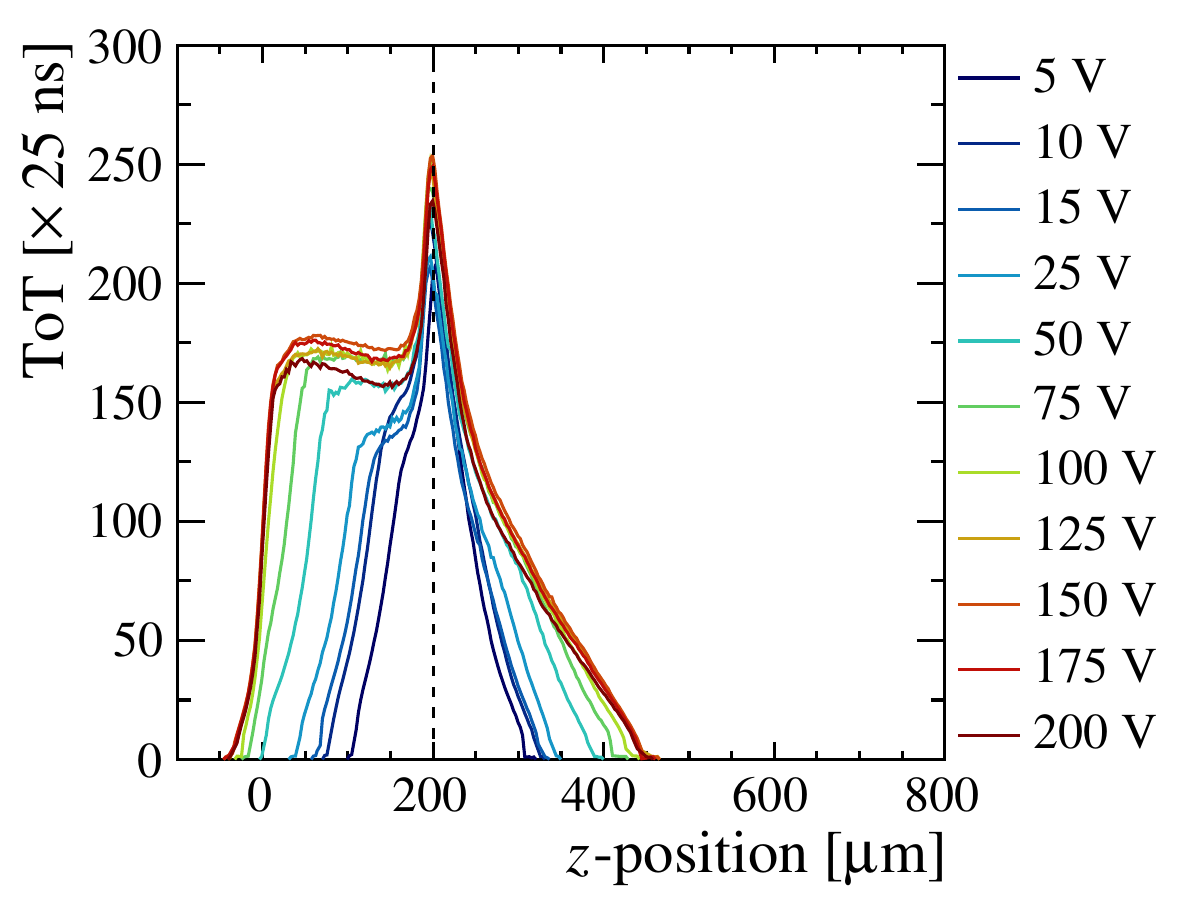}}\hspace{1mm}
  {\includegraphics[width=0.49\textwidth]{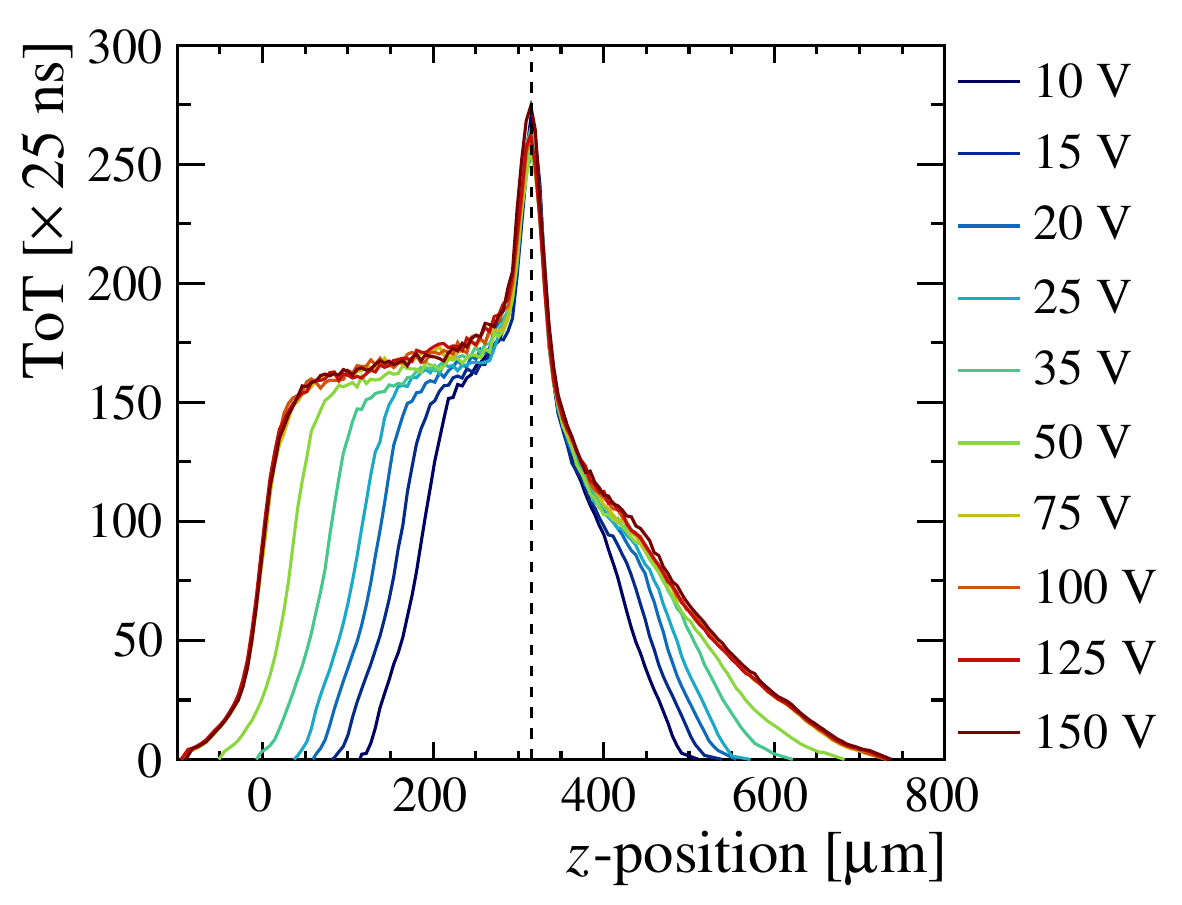}}
  \caption{ Time-over-Threshold (ToT) as a function of the voxel $z$-position for different bias voltages for the 200\mum \np sensor (left) and the 300\mum \pn sensor (right). The ToT is proportional to the amount of charge collected. If the bias voltage is below the depletion voltage the effective thickness of the sensor decreases. The vertical dashed line indicates the physical thickness of the sensor.
  }
  \label{fig:biasScanToT}
\end{figure}

Because the longitudinal TPA voxel length is larger than its transversal width, the spatial resolution that can be achieved perpendicular to the normal of the bulk is better. Nonetheless the resolution of the voxel is good enough to also resolve structures along the longitudinal direction. 
A scan along the longitudinal direction is shown in \fig\ref{fig:biasScanToT} (left:~200\mum~\np, right:~300\mum~\pn). 
The 0.7 NA objective was used for the 200\mum~\np sensor and the 0.5 NA objective was used for the 300\mum~\pn sensor. 
This \fig shows the ToT as a function of position along the beam direction ($z$-axis corrected for the length scaling in silicon), for different bias voltages. 
The different height of the ToT plateaus is due to a different starting phase of the TPA fluctuations, resulting in a different scale factor between the two scans. 
In \fig\ref{fig:biasScanToT} the 0\mum position corresponds to the back of the sensor and the 200\mum position (or 315\mum) corresponds to the front of the sensor. The number of charge carriers liberated in the sensor for different $z$-positions, assuming negligible reflection at the silicon-air interface, is given by \eq\ref{eq:Ntpa}~\cite{wiehe2020development}, where $d$ is the thickness of the sensor. 
\begin{equation}
    N_{tpa}\left( z \right) = \frac{E_p^2 n \beta_2 \sqrt{\ln{4}}}{4 c \hbar \pi^{3/2} \tau} \left[ \arctan\left( \frac{d-z}{z_0} \right) + \arctan\left( \frac{z}{z_0} \right) \right].
    \label{eq:Ntpa}
\end{equation}

Since for hybrid sensors the reflection at the front of the sensor is not negligible due to the presence of the metallisation, only the left side of these scans (below \SI{185}{\micro\meter}) are fitted using \eq\ref{eq:Ntpa} to retrieve the effective position of the back side, while the position of the front side is assumed to be at the centre of the peak (caused by reflection) and is approximated by a Gaussian distribution. From this, the effective thickness of the \np (\pn) sensor is determined to be \SI{201(1)}{\micro\meter} (\SI{315(1)}{\micro\meter}), where the quoted uncertainties are conservative estimates, expected to be dominated by contributions from the the motion stages. These values are consistent with the thicknesses stated by the manufacturers. 

The $z$-scans at different bias voltages also show the depletion depth growing with higher bias voltages. Around \SI{100}{\volt} is needed to collect all charge from the back side of the 200\mum \np sensor, which is not consistent with the depletion voltage of \SI{120}{\volt} determined using charge collection efficiency~\cite{geertsema2021charge} and grazing angle~\cite{dall2021temporal} data. For the 300\mum \pn sensor, a bias voltage of around \SI{75}{\volt} is needed to completely deplete the sensor. 
The lower depletion voltage found using the TPA setup is due to the collection of charge from the non-depleted region by means of diffusion, which in the charge collection and grazing angle analysis was not (or only partially) possible because of charge being liberated along the entire thickness of the sensor (line-like) instead of a point-like volume. In those analyses the random time interval between hits in some cases is smaller than the time it takes a single hit to cross the threshold. Therefore it is not possible to precisely assign a hit with a large drift time to a trigger.

\subsection{Time resolution studies}
\label{sec:timeresstudies}

The fast nature of the optical pulses of the TPA setup makes it possible to study the temporal characteristics of the sensors as a function of the 3D position within the sensor. In this \sect the time resolution as well as the Time-to-Threshold (TtT) of the two sensors under study are discussed.

Each hit is assigned to a corresponding trigger from the SPIDR TDC within a time window of \SI{1}{\micro\second}. Since the laser pulses are spaced by \SI{126}{\micro\second}, this windows leaves sufficient time not to associate triggers from different pulses. This selection ensures that the hits are generated by laser pulses, and not by noise or background radiation. The TtT is determined after this selection, and is defined as the time difference between the trigger and the hit. The time resolution~($\sigma_t$) is defined as the RMS of this TtT distribution. The RMS is determined for each position by fitting the TtT distribution with a Gaussian function.

\begin{figure}[t!]
  \centering
  {\includegraphics[width=0.49\textwidth]{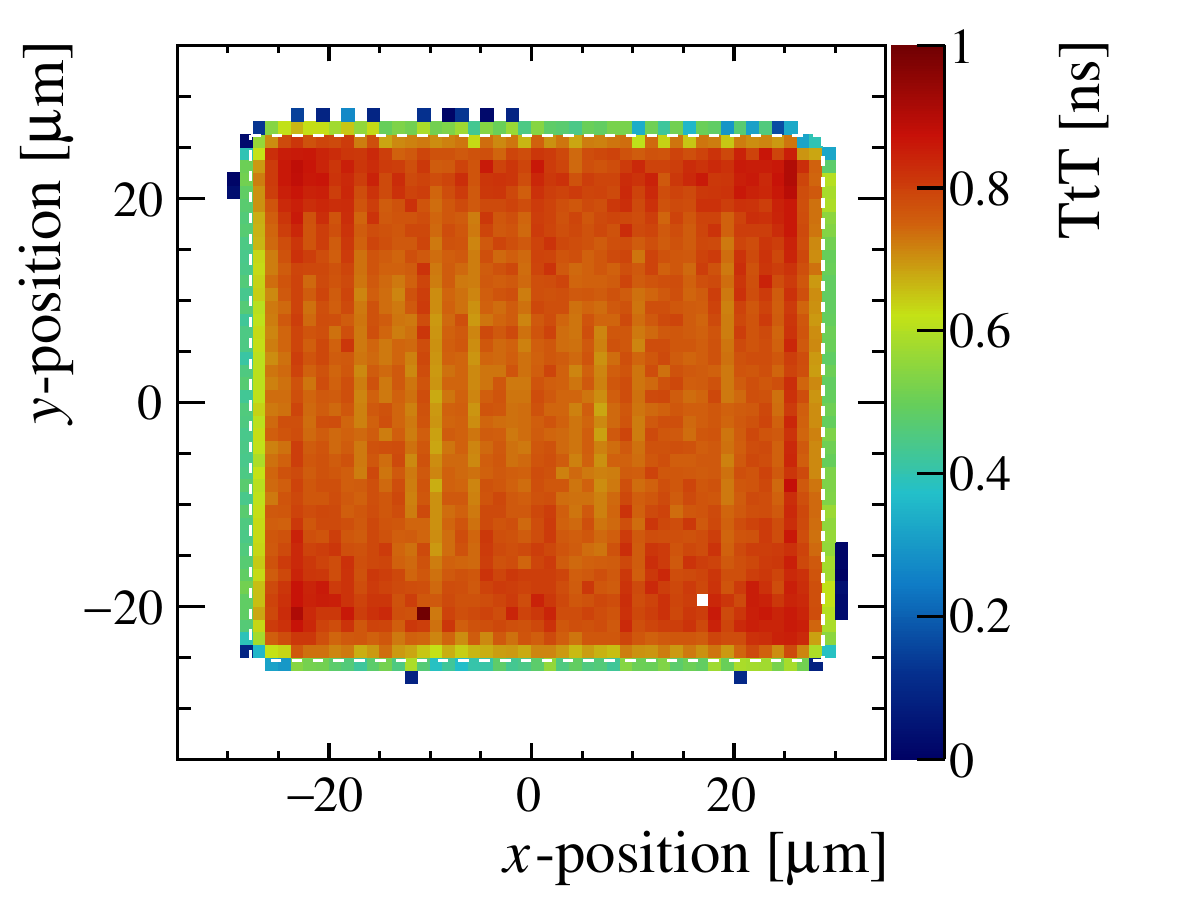}} \hspace{1mm}
  {\includegraphics[width=0.49\textwidth]{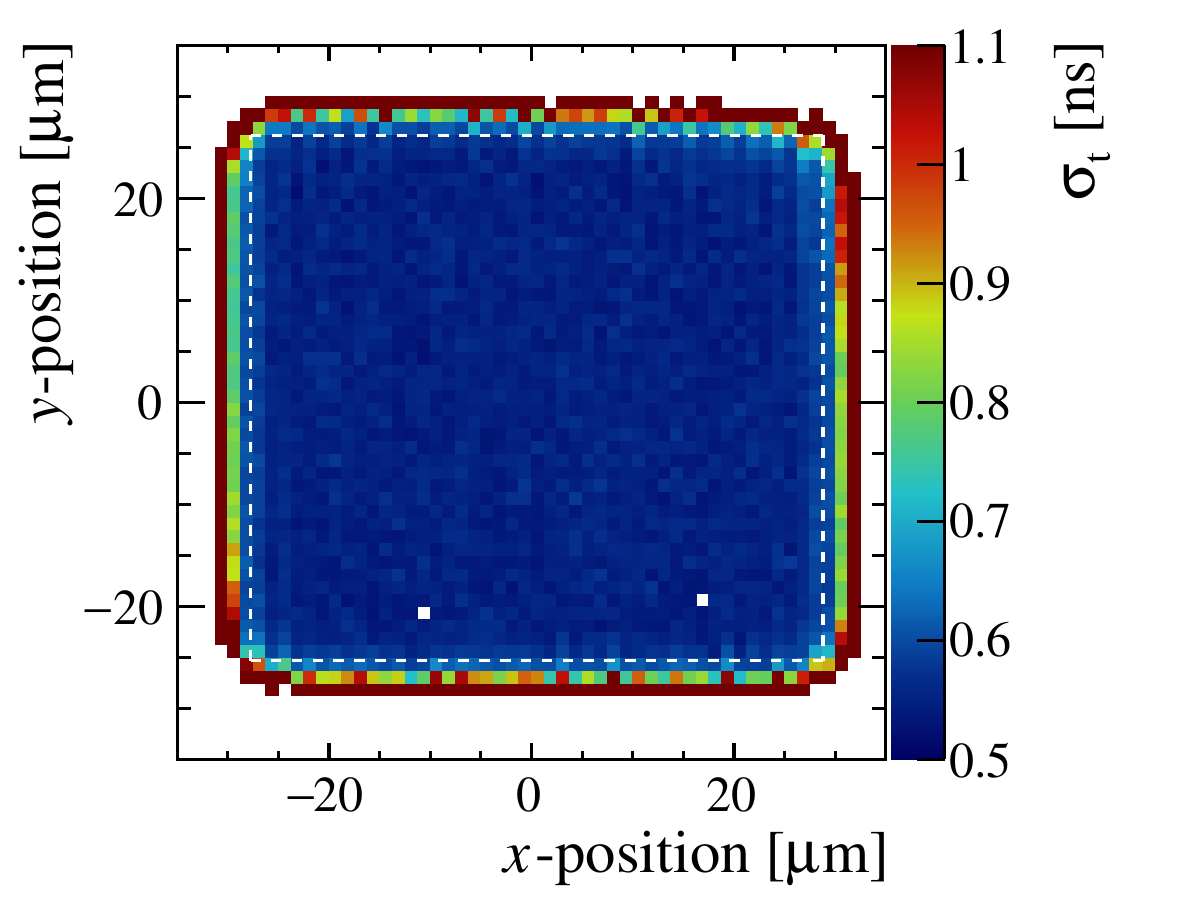}}
  \caption{Time-to-Threshold, (TtT), (left) and time resolution (right) for different positions in and around a single pixel in the \np sensor. The dashed white lines indicate the outline of the pixel. The one (two) white dot(s) that are visible in the left (right) plot indicate positions where the fit did not converge. }
  \label{fig:temporalTimeRes}
\end{figure}

The standard deviation and the average of the TtT distributions for the area scan in the middle of the bulk of the pixel are presented in \fig\ref{fig:temporalTimeRes} (same data set as \Fig\ref{fig:ToTscans}). The white dashed line in these figures indicates the outline of the pixel. The TtT (left) shows a uniform behaviour over the complete pixel. Slight variations in the TtT can be observed near the centre of the pixel. These, however, are due to intensity fluctuations in combination with the residual timewalk because of the imperfect timewalk correction.
For low ToT values the TtT is consistently below (or above) the parameterisation, which causes these ToT values to be systematically under (over) corrected.
Therefore a variation in ToT results in a TtT spread, which can be observed as vertical bands in \fig\ref{fig:temporalTimeRes}~(left). 
However, the differences at the corners and the pixel edges are sensor effects. Towards the corner of the pixel it can be seen that the TtT becomes longer. This is also observed when the TPA voxel is outside of the pixel under test. 
This is due to an increased travel time of the charge carriers to reach the implant, leading to an increased lateral diffusion, and subsequently to more charge sharing.
As mentioned before, for low ToT hits the timewalk correction (\sect\ref{sec:Timepix3ASIC}) performance is less optimal which results in the systematic under- or over-correction of the TtT.
An improved timewalk calibration is currently planned, and requires additional work on the Timepix3 data which is not the main scope of this paper.

The time resolution, shown in \fig\ref{fig:temporalTimeRes}~(right), shows similar behaviour as the TtT. The resolution is uniform over almost the entire pixel, with the exception of the four corners, where the charge is shared among four pixels. 
This is similar to the uniform behaviour that is observed using a charged particle beam~\cite{dall2021temporal}, although the time resolution found in this paper is significantly better than that observed with the beam data. 
The difference is attributed to two main effects. 
Firstly, the time resolution measured using lasers is expected to be better due to smaller charge fluctuations compared to the energy loss straggling of a charged particle. 
Secondly, the measurements presented in~\cite{dall2021temporal} combines data of all pixels together and this convolutes the matrix fluctuations and clock distribution effects. This degrades the time resolution compared to that of a single pixel, see~\cite{heijhoff2020timing}.

\begin{figure}[t!]
  \centering
  {\includegraphics[width=0.49\textwidth]{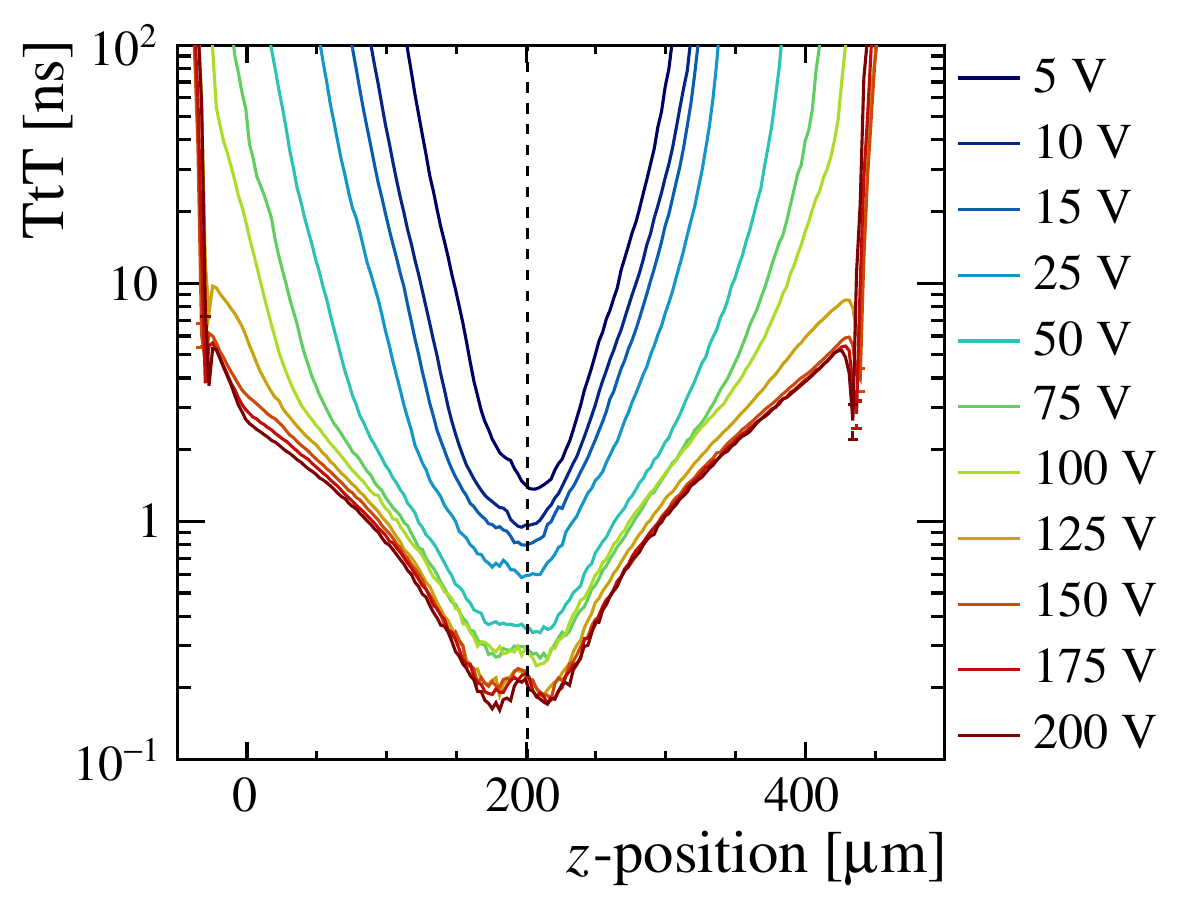}} \hspace{1mm}
  {\includegraphics[width=0.49\textwidth]{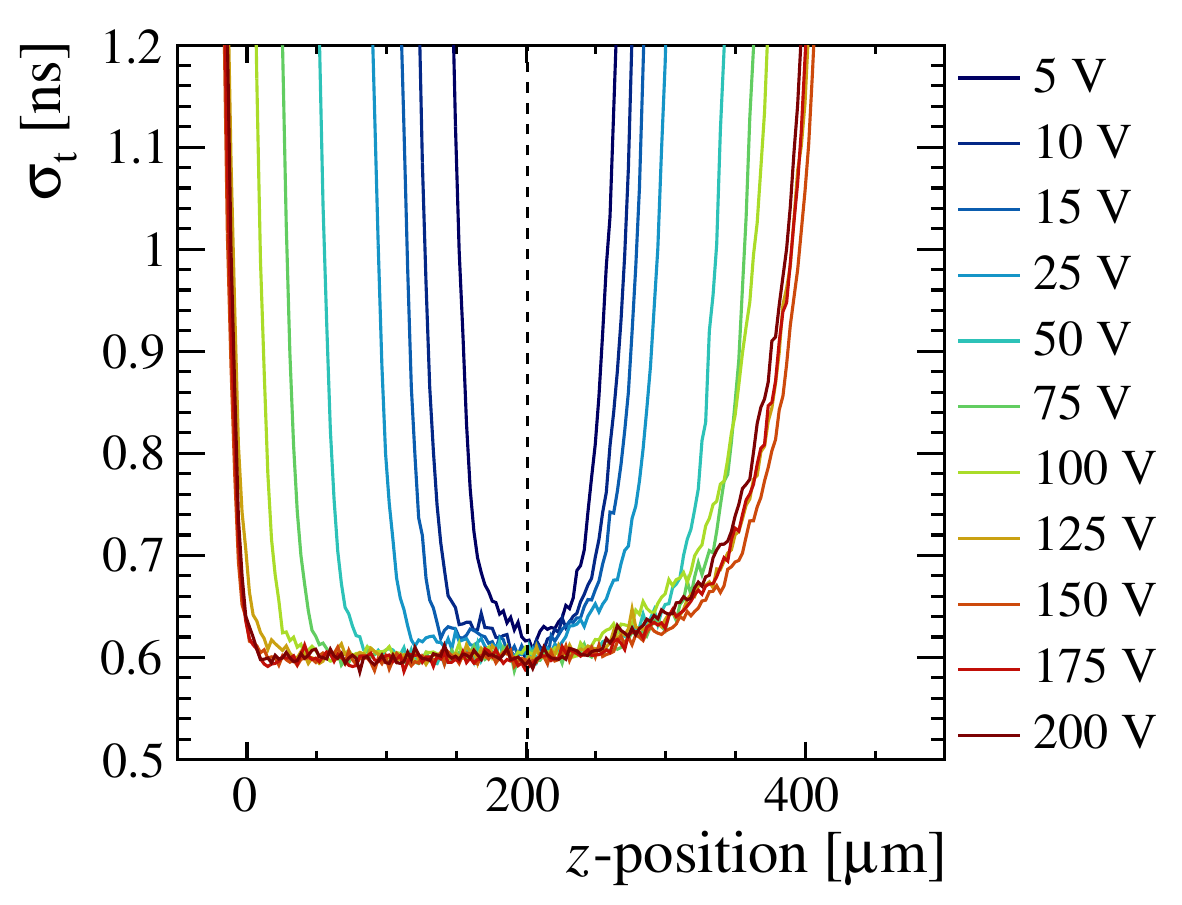}}\\
  {\includegraphics[width=0.49\textwidth]{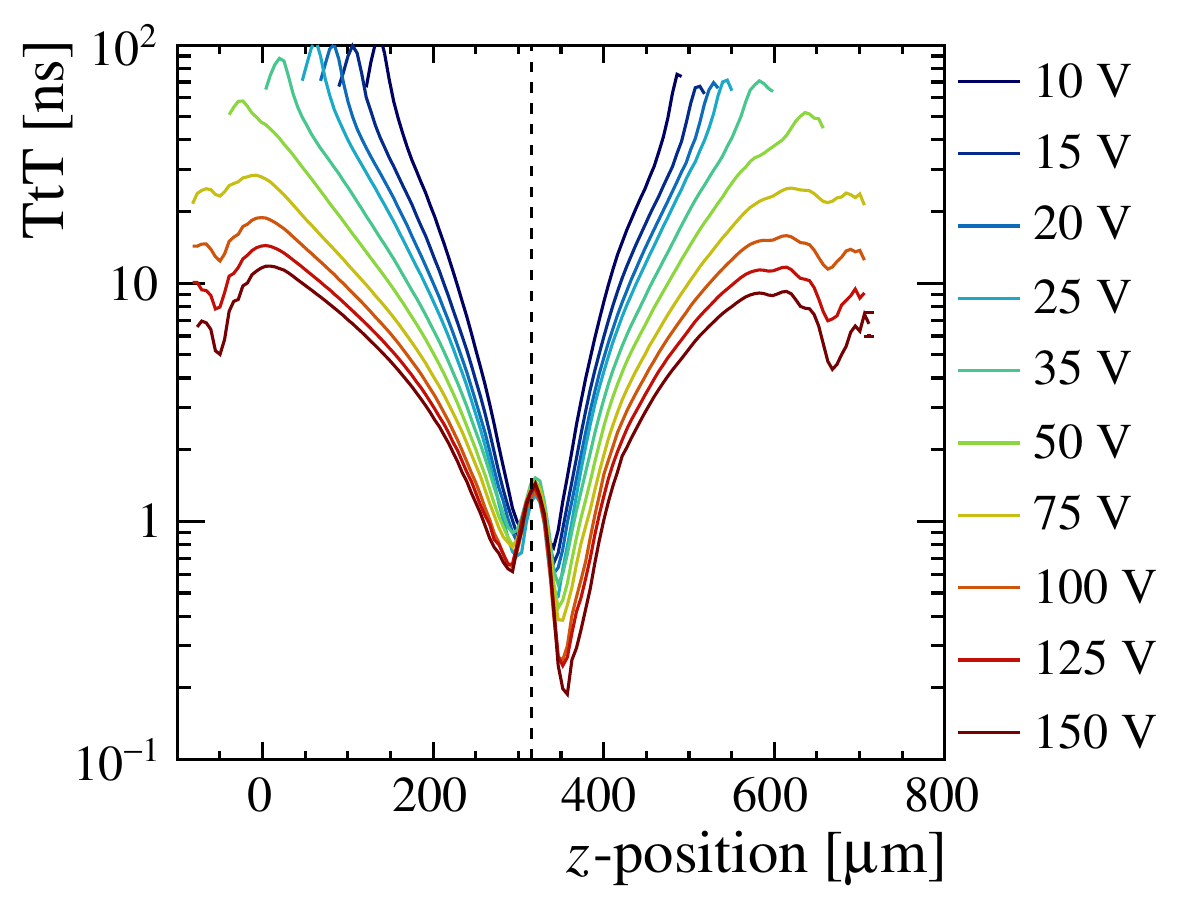}} \hspace{1mm}
  {\includegraphics[width=0.49\textwidth]{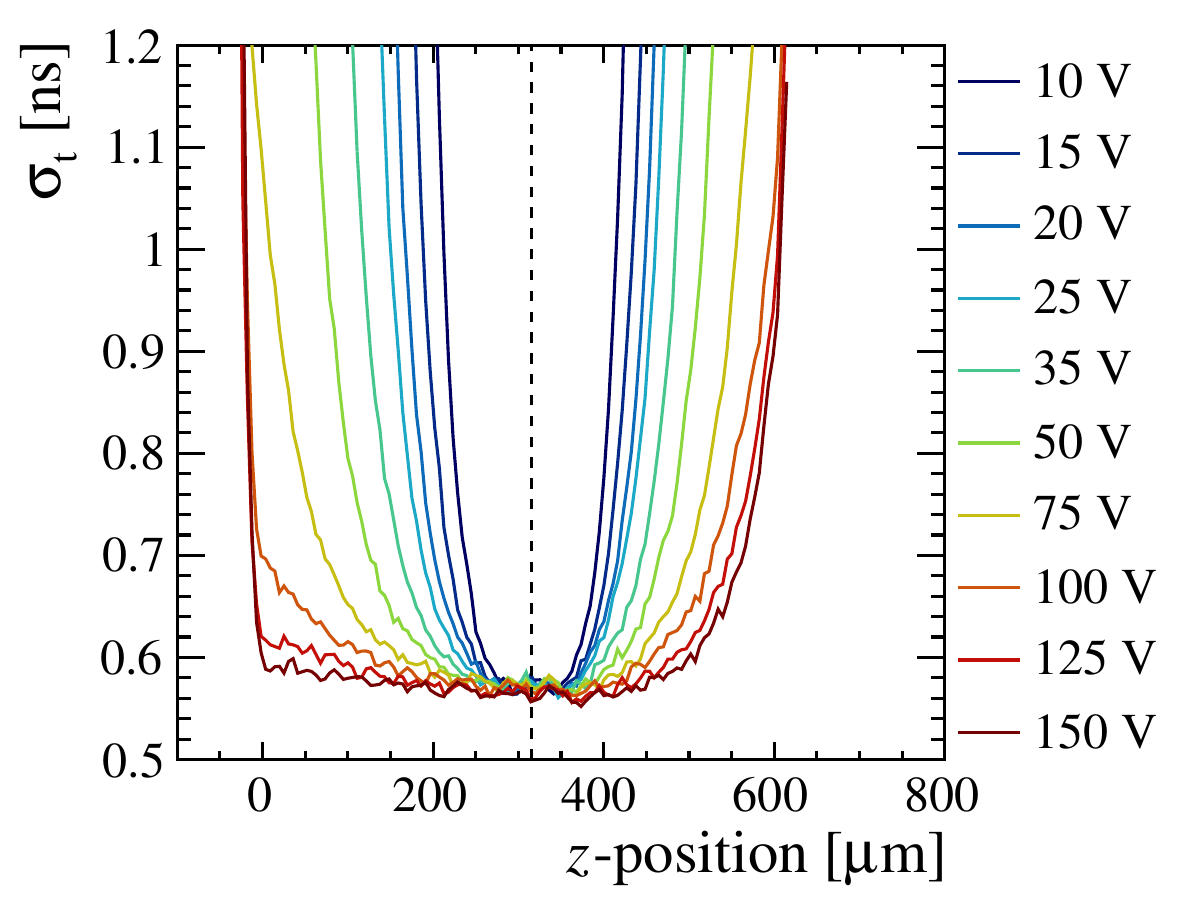}}
  \caption{Time-to-Threshold, (TtT), (left) and time resolution (right) as a function of depth within the sensors at different bias voltages. The top (bottom) row shows the results of the 200\mum \np sensor (300\mum \pn sensor). All plots have error bars indicating the statistical uncertainties from the fit, however most are too small to see.}
  \label{fig:timingZscan}
\end{figure}

The TPA voxel can also be moved throughout the depth ($z$-direction) of the sensor by moving the $z$-stage. During these scans, the TPA voxel is positioned in the centre of the pixel. \Fig\ref{fig:timingZscan} shows the TtT and the time resolution as a function of the depth in the bulk for the 200\mum \np sensor (top), and the 300\mum \pn sensor (bottom). The $z$-position for these scans is corrected for the difference in refractive index as discussed in \sect\ref{sec:TPA}. Below, the 200\mum \np sensor will mainly be discussed since most observations are consistent between the two sensors. As mentioned in \sect\ref{sec:ChargeMeasurements}, the 200\mum \np sensor is actually measured to be \SI{201}{\micro\meter} thick. However, in \fig\ref{fig:timingZscan} the depth is scanned from approximately the back of the sensor, to about \SI{250}{\micro\meter} beyond the front of the sensor (\SI{0}{\micro\meter} to \SI{450}{\micro\meter} in the plot). The signal observed from \SI{200}{\micro\meter} to \SI{450}{\micro\meter} is due to internal reflection of the laser light at the front and back interfaces.

\Fig\ref{fig:timingZscan}~(left) shows the TtT as a function of the voxel depth in the bulk, where different colored curves denote different bias voltages. The \figNoSpace~contains error bars indicating the statistical uncertainty of the fit. However, for most data points the error bars are too small to distinguish in the figure. The TtT is, as expected, larger for positions far away from the implant, and low for positions close to the implant. The bias voltage also influences the TtT significantly, since the magnitude of the induced current and hence the slope of the integrated current, is lower for low drift velocities due to the lower electric field at low bias voltages. The small increase in time resolution present when the TPA voxel is first reflected (around \SI{215}{\micro\meter}) is not yet understood. Also a small decrease in the TtT around \SI{0}{\micro\meter} and \SI{450}{\micro\meter} is observed which is due to a systematic under- or over-correction of the timewalk as discussed in \sect\ref{sec:Timepix3ASIC}. 

The time resolution throughout the depth of the sensor for different bias voltages is shown on the right side of \fig\ref{fig:timingZscan}. 
 For the highest bias voltage (\SI{200}{\volt}), the resolution remains constant at about \SI{600}{\pico\second} (\SI{560}{\pico\second} for the 300\mum sensor at \SI{150}{\volt}) for almost all positions probed within the sensor. 
 The resolution degrades when the voxel is placed at the back side of the sensor (low $z$-values) caused by a loss in signal magnitude due to a partial depletion for lower bias voltages, and thus of the TPA voxel is outside of the depleted region. 
 The resolution is also observed to be approximately constant for the different bias voltages, with degradations mainly caused by the loss in signal magnitude. This is observed when the signal decreases after internal reflection between \SI{200}{\micro\meter} and \SI{450}{\micro\meter}. 
 The main difference between the time resolution of the two assemblies is caused by pixel-to-pixel variations giving the ASIC contribution of the two different pixels slightly different time resolutions, and not by the resolution of the sensors itself. Since the resolution of the sensors is significantly smaller than of the ASIC, a small difference in the sensor contributions is not expected to be observed.

As discussed in \sect\ref{sec:siliconSensors}, the measured time resolution is the effective results of the Timepix3 TDC, time resolution of the analogue front-end of Timepix3, SPIDR TDC, and the time resolution of the silicon sensor. These components, excluding the sensor resolution, combine to a time resolution of \SI{516}{\pico\second}. Therefore, from the overall resolution of \SI{600}{\pico\second} (\SI{560}{\pico\second}), about \SI{306}{\pico\second} (\SI{218}{\pico\second}) is attributed to the sensor for the \np (\pn) type. The resolution found for these two sensors is still significantly higher compared to resolutions found for similar planar sensors. For a planar 200\mum sensor a resolution of \SI{127.1}{\pico\second} (including ASIC contributions) is reported previously~\cite{rinella2019na62}. This may indicate that the time resolution of the characterised sensors is better, and that currently we are limited by the pixel-to-pixel variations in the ASIC which dominates the fluctuations on the time resolution.

%% file: 05_Conclusion.tex
\section{Conclusion and outlook}
\label{sec:Conclusion}

TPA techniques provide a method that enables the characterisation of the complete detector volume, which is in contrast to using SPA techniques. The strong optical focusing needed for TPA also leads to small voxel sizes where charge is liberated, leading to a good spatial resolution with which the detectors can be studied. 

An initial investigation of the capabilities of the newly commissioned TPA setup at Nikhef is presented. Characterisation studies of this system show an output power variation below 1\%. However, the TPA signal presents an instability measured to be about 1.6\%, which is found to be linked to a variation of the pulse time width over time. This instability is corrected offline using a TPA reference signal generated on a silicon diode with 85\% of the beam power. The system is equipped with a precise timing trigger which is shown to have a resolution better than~\SI{30.4}{\pico\second}. This resolution is dominated by its reference detector, and therefore is, most likely, significantly better.

Two sensors connected to Timepix3 ASICs have been tested with the TPA system. The charge collection and temporal behaviour were presented. The charge collection is found to be uniform over the pixel, and scans along the beam direction indicate a thickness of the sensor that is in agreement with the thickness stated by the manufacturer. Light is reflected by the metallisation at the front side of the sensor, yielding a small region with increased intensity and thus a larger charge collection. This effect proved useful in imaging the shape and height differences of the front side metallisation. 

The combined time resolution of the sensor and the ASIC is found to be dominated by the ASIC contributions, and is measured to be \SI{600}{\pico\second} and \SI{560}{\pico\second} for the 200\mum \np and 300\mum \pn sensors, respectively. An overall degradation of the time resolution and the TtT is observed at lower bias voltages due to lower charge collection and lower drift velocities. Studies with sensors and ASICs with better timing resolution are expected to be performed soon and will utilise the full potential of this TPA system in space and time.

%% file: 06_acknowledgements.tex
\section*{Acknowledgments}

We express our sincere gratitude to Oscar van Petten for his vital support in constructing the mechanics for the TPA setup. 
This research was funded by the Dutch Research Council (NWO) with co-financing from the Dutch public-private partnership (PPP) allowance for research and development in the top sector High Tech Systems and Materials (HTSM).